**Hydrogen-Induced Sign Reversal in Magnetic Hysteresis Evolution of CoPd Alloys and Co/Pd Multilayers**

S. S. Das[1,*] and A. Gerber[2,*]

[1]Institute of Industrial Science, The University of Tokyo, 4-6-1 Komaba, Meguro-ku, Tokyo 153-8505, Japan

[2]Raymond and Beverly Sackler Faculty of Exact Sciences, School of Physics and Astronomy, Tel Aviv University, Ramat Aviv, 69978 Tel Aviv, Israel

**Abstract:** Hydrogen absorption in magnetic thin film nanostructures can modulate their electronic, magnetic, and transport properties by modifying the electronic structure and lattice strain. However, the influence of composition and nanostructuring on these two competing effects is not well understood. In this study, we systematically investigate the evolution of hydrogen-induced magnetic hysteresis in three complementary CoPd systems: $Co_xPd_{100-x}$ alloys, $[Co(0.1\ nm)/Pd(d)]_{15}$, and $[Co(0.2\ nm)/Pd(d)]_{15}$ multilayers, using extraordinary Hall effect characterizations (EHE) in air and under a 4% $H_2/N_2$ mixture. We show that hydrogen-induced magnetic response in the CoPd system is not universal, but instead exhibits a strong dependence on composition and layer thickness. This is governed by the interplay between Pd-related electronic effects and magnetoelastic anisotropy. In Pd-rich CoPd alloys and in Co(0.2 nm)/Pd multilayers, hydrogen initially contracts the hysteresis loops at low Co fractions, followed by a transition to loop expansion above an effective Co concentration of x ~ 40 %. This contrasting behavior is the result of the interplay between the electronic suppression of the Pd-induced magnetization due to the filling of the Pd-4d band, and the hydrogen-driven, anisotropic strain that strengthens the magnetoelastic anisotropy in Co-rich samples. In contrast, in ultrathin $[Co(0.1\ nm)/Pd(d)]_{15}$ multilayers,

*Corresponding authors: ssdas@iis.u-tokyo.ac.jp (S. S. Das) and gerber@tauex.tau.ac.il (A. Gerber)

hydrogen induces a weak, non-monotonic but generally expanding loop behaviour across the entire composition range (x = 15–60 %), indicating a dominant role of interfacial magnetic connectivity and strain-mediated magnetoelastic anisotropy in the ultrathin limit. Furthermore, our data reveals a hydrogen-induced reversal of the EHE loop polarity near the crossover regime, reflecting a change in the dominant EHE scattering mechanisms, thus providing an additional degree of magnetic tunability by hydrogen. These results demonstrate that hydrogen can selectively tune the magnetism of CoPd nanostructures via composition-controlled electronic and magnetoelastic effects, offering insights into the design and development of hydrogen-responsive spintronic and sensing devices.

**Graphical Abstract:**

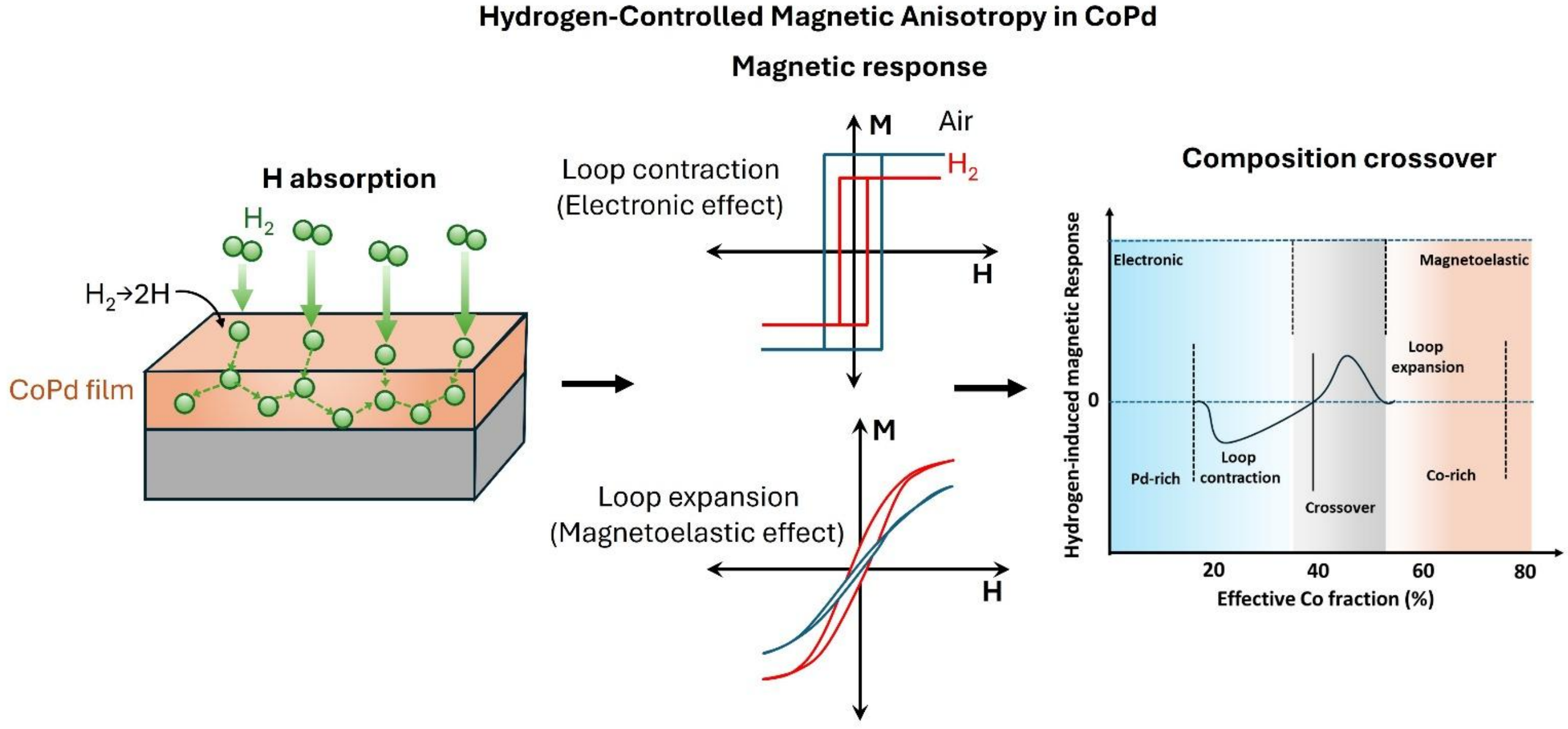

## I. Introduction

Hydrogen, being the smallest and lightest element, can easily penetrate into the transition metal lattice, modifying its electronic, magnetic, and transport properties [1,2]. Palladium and Pd-based alloys are particularly well known for their exceptional hydrogen catalytic properties and remarkable hydrogen-responsive physical properties [2–4]. Molecular hydrogen, upon interaction with Pd surface, undergoes self-dissociation into atomic hydrogen, which later diffuses into the metal lattice and occupies the interstitial sites [1,4–6]. In the case of FCC-Pd, the octahedral (O) site, due to its larger free volume, is the most stable site for H occupancy relative to the tetrahedral (T) site [7,8]. Hydrogen occupancy at O-sites induces two important effects: (i) modification of Pd 4d-band occupancy and associated electronic structure [1–4] and (ii) a significant lattice expansion due to interstitial occupancy [3–5]. Together, these electronic and elastic effects in Pd-based magnetic materials modify exchange interactions, magnetoelastic energies, magnetic anisotropy, and magnetotransport properties [5,9–14], making them an attractive choice for exploring hydrogen-controlled spin physics and developing hydrogen-responsive magnetic devices for emerging hydrogen technology.

Since the demonstration of EHE-based magnetic hydrogen sensing in CoPd systems, the field of magnetic hydrogen sensing has evolved significantly. Recent studies have explored several magnetic-based detection approaches, including the magneto-optical Kerr effect, ferromagnetic resonance, and magnetoresistance [10,13,15–19]. These studies demonstrate that hydrogen absorption can modulate magnetic anisotropy, spin-reorientation transitions, magnetotransport, and spin–orbit-coupled magnetic responses in Pd-based magnetic alloys and multilayers. In particular, Co–Pd and Fe–Pd systems have emerged as model platforms for exploring hydrogen-controlled magnetism and developing hydrogen-sensitive spintronic devices [16,18]. Although these studies highlight the hydrogen-induced modification of magnetization loops, a systematic understanding of how such loop evolution depends on CoPd composition and Co/Pd multilayer nanostructures remains limited.

CoPd alloys and ultrathin Co/Pd multilayers constitute two structurally different classes of Pd-based magnetic material that exhibit perpendicular anisotropy (PMA) arising from different microscopic mechanisms [11,14,20,21]. The origin of PMA in Pd-rich CoPd alloys arises from a combination of electronic and magnetoelastic contributions [20,22,23]. Hybridization between Co-3d and Pd-4d orbitals, amplified by the strong spin–orbit coupling of Pd, reshapes the d-electron cloud in an anisotropic manner that lowers the energy for out-of-plane magnetization [22,24–26]. In addition, the magnetoelastic contribution further influences the magnetic anisotropy and depends strongly on the (111)-oriented magnetostriction coefficient $\lambda_{111}$ [21]. Importantly, $\lambda_{111}$ varies systematically with alloy composition: Pd-rich alloys exhibit a negative magnetostriction coefficient with maximum at 15 at. % Co, whereas Co-rich alloys show positive or weakly positive values, with a crossover typically occurring around 40–50 at. % Co [27,28]. This composition-dependent sign of change of $\lambda_{111}$ leads to qualitatively different responses to the internally generated stress. The role of substrate-induced and growth-induced strain is also crucial in determining PMA in thin CoPd films [21,23,29]. On amorphous glass substrates, the absence of a well-defined crystallographic relationship restricts the appearance of a coherent epitaxial strain, thus reducing the magnetoelastic contribution to the PMA [23,30,31].

In contrast, ultrathin Co/Pd multilayers exhibit PMA due to interfacial Co-Pd hybridization, spin-orbit coupling, and strain-related effects [20,31]. At the Co/Pd interface, broken symmetry and Co–Pd hybridization reshape the Co 3d orbitals ($d_{xz}$, $d_{yz}$, and $d_{z^2}$) leading to a larger out-of-plane orbital moment than in-plane [20,23,31,32]. Since the spin and orbital moments are strongly coupled through spin–orbit interaction, this orbital moment anisotropy directly induces spin alignment in the out-of-plane direction [25,33]. As a result, Co/Pd multilayers show robust PMA at ultrathin Co thicknesses.

This interfacial orbital-moment anisotropy and the role of Pd proximity effects have also been supported by XMCD and first-principles calculations [26]. Previous Extended X-ray Absorption Fine Structure (EXAFS) studies have shown that the local Co-Pd atomic arrangement differs between CoPd alloys and Co/Pd multilayers, and that CoPd hybridization near Pd neighbors plays an important role in PMA formation [34].

Hydrogen absorption by Pd and Pd-based systems simultaneously changes their electronic structure as well as introduces strain effects [1,4,13]. Hydrogen donates its 1s electron to the Pd 4d-band, increasing its filling and altering the spin polarization, which in turn modifies the exchange interactions [1,35]. DFT calculations on the Pd-H interaction have shown that hydrogen absorption results in strong Pd 4d–H 1s bonding, which modifies the Pd d-state filling and strongly reduces the density of states near the Fermi level [36]. In Co/Pd-based magnetic systems, first-principles studies have shown that hydrogen can modify the local hybridization between H 1s states and metal *d* orbitals (Pd-4d, Co-3d), thereby changing the magnetic anisotropy depending on hydrogen position and concentration [32]. Akamaru et al. further showed experimentally and by first-principles calculations that hydrogenation reduces the Curie temperature and magnetic moment of Pd–Co alloys by weakening Co spin polarization and the Pd spin polarization induced by Co [40]. In addition, in operando polarized neutron reflectometry/ferromagnetic resonance (PNR/FMR) measurements together with DFT calculations have shown that hydrogen absorption weakens the PMA associated with the Co/Pd interface, and alters the spin polarization near the Fermi level [13]. Furthermore, FMR studies on Pd/Co layered films by Lueng et al. have compared separate measurements under hydrogen exposure and externally applied strain, suggesting that hydrogen-modified PMA is mainly electronic in origin, while the magnetoelastic contribution is small with opposite sign [37]. These studies support that hydrogen-induced electronic modification can weaken Pd-derived magnetism and PMA. Apart from electronic modification, the H occupancy at the interstitial sites also causes anisotropic lattice expansion in the case of thin films clamped on a hard substrate [3,38,39]. This generates a large H-induced in-plane compressive stress that affects the magnetoelastic anisotropy energy in magnetostrictive systems like Co-Pd [23,40]. This combined electronic and magnetoelastic effect decides the overall hydrogen-induced magnetic response of the system. Several experimental studies have also evidenced a decrease in perpendicular magnetic anisotropy upon hydrogen absorption. However, hydrogen response in magnetic systems is not universal, it can vary depending on composition, interface structure, layer thickness, strain, and microstructure. Previous studies on the hydrogen-induced magnetic responses have reported either hysteresis loop contraction or expansion of specific compositions or individual multilayer structures, without addressing a comprehensive picture of their composition or thickness dependence and the origin of these contrasting changes [11,12,41]. The hydrogen-induced response in CoPd system can be tuned by its composition, microstructure, and interface modification [11,20]. Despite several studies, a systematic understanding of the hydrogen induced magnetic hysteresis changes in CoPd systems (including CoPd alloys across the entire composition range and in ultrathin Co/Pd multilayers) is still lacking. Moreover, a comprehensive picture of the interplay between the hydrogen-induced electronic effect and magnetoelastic effects in controlling the overall magnetization hysteresis changes of the system is interesting both from fundamental and technological reasons.

In this work, we systematically studied the hydrogen-induced evolution of magnetic hysteresis loops in CoPd alloys and Co/Pd multilayers, using extraordinary Hall effect measurements under air and $H_2/N_2$ atmosphere at room temperature. By varying the stoichiometry of CoPd alloys and respective layer thicknesses in ultrathin Co/Pd multilayers, we achieve a controlled transition from Pd-dominated magnetism (low-Co alloys, Co(0.2 nm) multilayers) to Co-dominated magnetism (high-Co alloys, ultrathin Co(0.1nm) multilayers). Our results reveal a composition-dependent reversal in hydrogen-induced hysteresis response, allowing us to disentangle the competing roles of Pd-related electronic effects and magnetoelastic anisotropy. These studies offer insights for the design and development of hydrogen-responsive spintronic devices by varying alloy composition and nanoscale structural design.

## II. Experimental

Polycrystalline $Co_xPd_{100-x}$ alloy films were deposited simultaneously onto room-temperature glass and GaAs(001) single-crystal substrates, while ultrathin $[Co/Pd]_{15}$ multilayer samples were grown on glass substrates only, using the RF magnetron sputtering technique in a high vacuum chamber of base vacuum $5 \times 10^{-7}$ Torr. The sputtering setup is equipped with three sputtering targets and programmable shutters, enabling sequential deposition for multilayer films. Sputtering was carried out in Ar at a working pressure of $P_{Ar} = 3 \times 10^{-3}$ Torr. The stoichiometry of the CoPd alloys was controlled by adjusting the RF power applied to the individual Co and Pd targets. Co and Pd form a uniform solid solution across the full composition range, even at room temperature [42]. For the multilayer samples, RF powers of 15 W and 10 W were applied to the Co and Pd targets, respectively. The corresponding deposition rates were ~0.014 nm/s for Co and ~0.043 nm/s for Pd. Two series of multilayers were fabricated: (i) $[Co(0.1\ nm)/Pd(d)]_{15}$ and (ii) $[Co(0.2\ nm)/Pd(d)]_{15}$, where *d* is the Pd layer thickness, which was varied in the range 0.05–2 nm. For these multilayers, the effective Co concentration was calculated from the nominal Co and Pd layer thicknesses as $x_{\text{eff}} = \frac{t_{\text{Co}}}{t_{\text{Co}}+t_{\text{Pd}}} \times 100\%$, where $t_{\text{Co}}$and $t_{\text{Pd}}$ are the nominal Co and Pd thicknesses in each bilayer. This value is used as a nominal composition parameter for comparing the multilayer series with CoPd alloys and should not be interpreted as a direct measure of chemically sharp or continuous Co/Pd layers. The multilayer structure is formed through sequential deposition of Co and Pd layers. However, due to the ultrathin layer thickness, partial intermixing at the Co/Pd interfaces is expected, leading to alloy-like regions while preserving the overall multilayer architecture. Apart from this, since the nominal thickness of the Co layers used in the multilayers are only 0.1 nm and 0.2 nm, island-like Co coverage and interface roughness cannot be excluded.

Structural and microstructural characteristics of the films were investigated using high-resolution transmission electron microscopy (HRTEM, JEOL JEM-2100F) and selected area electron diffraction (SAED). For plan-view HRTEM analysis, the sample was sputtered directly onto the Cu grids. HRTEM images and SAED patterns were recorded to investigate the crystallinity and preferred texture of the films. The EHE measurements were performed at room temperature under ambient air and in a 4% $H_2/N_2$ mixture at atmospheric pressure. The measurement setup consisted of a GMV 3473 electromagnet, a Keithley 2400 source/meter, a Keithley 2001 multimeter, and an HP 3488A switch/control unit, controlled through a LabVIEW-based interface. The electrical contacts were made by bonding Al/Si wires in van der Pauw geometry, which was employed for both longitudinal resistance and EHE measurements. The fixed 4% $H_2/N_2$ condition was used to compare different samples under identical, near-saturated hydrogen exposure conditions. A representative concentration dependent EHE response of Glass/$Co_{36}Pd_{64}$(7 nm), shown in Fig. S5 confirms the systematic evolution of EHE response with $H_2$ concentration and also supports 4% $H_2/N_2$ as a near-saturated comparison condition. The sample was exposed to $H_2$ for ~45 min to reach saturation prior to the EHE measurement. The experimental sequence is detailed in Fig. S1 of the Supplementary file. The analysis is based on the first hydrogenation cycle, which enables us to record the intrinsic response of the pristine magnetic state. Subsequent cycles can introduce history-dependent effects due to residual hydrogen, defect stabilization or incomplete relaxation, which may alter the absolute values without changing the response mechanism.

## III. Results and discussions

### (A) Structural and microstructural characterization

The structural and microstructural properties of a representative 7 nm thick $Co_{32}Pd_{68}$ alloy film grown on Cu grids are shown in Fig. 1(a)–(b). The HRTEM micrograph reveals nanocrystalline morphology with randomly oriented grains, with an average grain size of ~ 8 nm. The relatively uniform grain distribution suggests homogeneous alloy formation across the film. The presence of grain boundaries indicates a fully polycrystalline nature of the grown film. The corresponding selected-area electron diffraction (SAED) pattern (Fig. 1(b)) exhibits concentric diffraction rings, confirming the polycrystalline nature of the film. The diffraction rings correspond to the (111), (200), (220), and (311) reflections, with the (111) reflection as the strongest. This suggests that the grains preferentially grow in (111) orientation, which is commonly observed in CoPd alloys [30,43]. From the diameter of the ring, the interplanar spacing ($d$) associated with the (111) reflection is estimated to be 0.23 nm, consistent with the reported values for fcc Pd-rich CoPd systems [44]. The presence of additional faint outer rings suggests the presence of higher-order reflections, indicating improved crystallinity of the film. Corresponding structural data for multilayers (not shown) also show a similar polycrystalline nature with (111) texture.

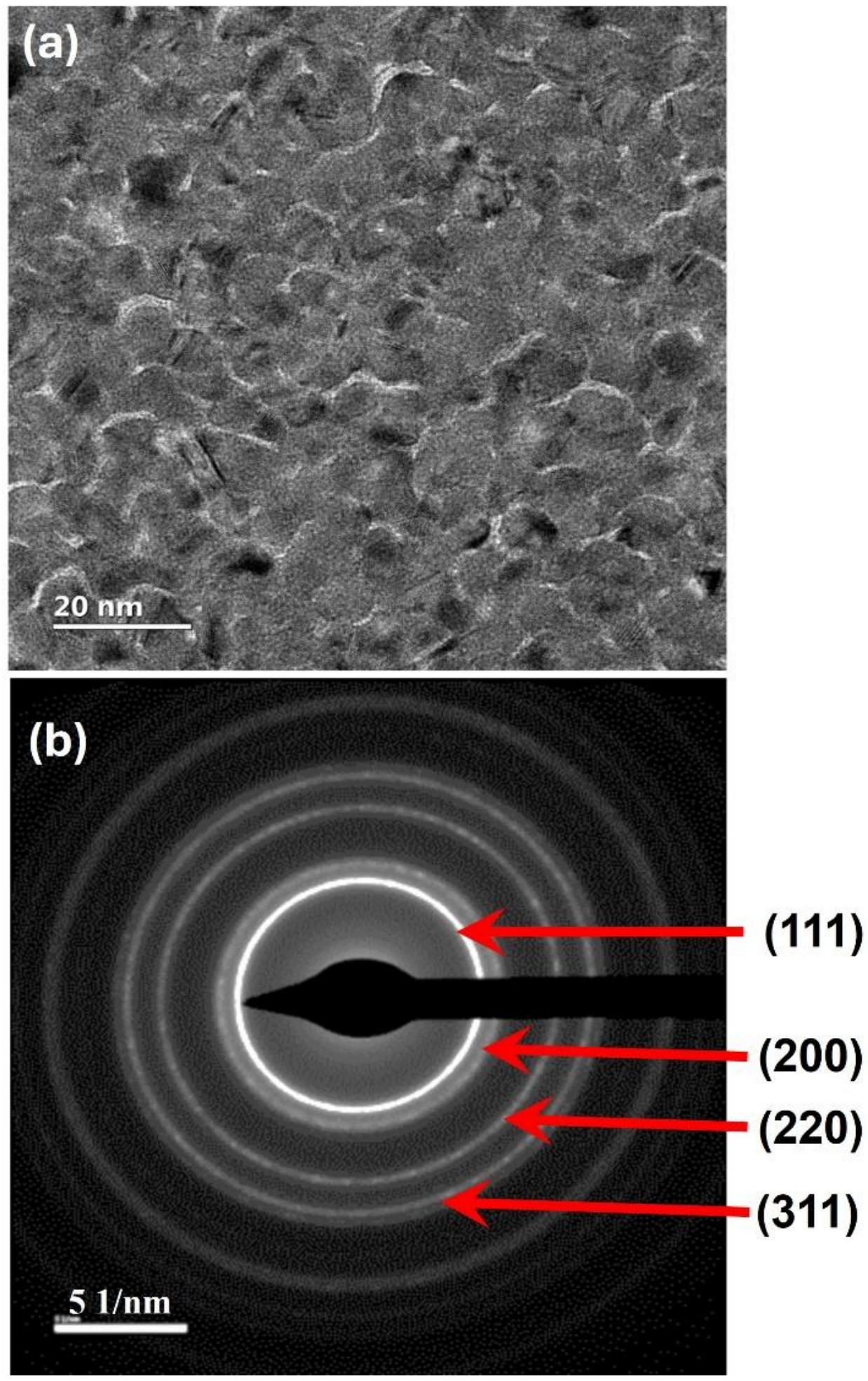

**Fig. 1** (a) HRTEM image of the $Co_{26}Pd_{74}$(7.5 nm) film showing nanocrystalline microstructure. (b) Corresponding SAED pattern exhibiting concentric diffraction rings confirming the polycrystalline fcc structure. The rings correspond to the (111), (200), (220), and (311) reflections. The outer rings correspond to higher-order reflections.

**(B) Hydrogen interaction and magnetic response in CoPd alloy films**

Figure 2 summarizes the sequence of hydrogen interaction with the CoPd film, from surface adsorption and dissociation to interstitial diffusion, together with the resulting impact on magnetic hysteresis. As illustrated in Figs. 2(a–c), hydrogen molecules ($H_2$) upon approaching the CoPd surface, first undergo physisorption and spontaneously dissociate into atomic hydrogen. The dissociative adsorption barrier in the case of Pd is negligible or very small compared with most metals, owing to its high catalytic efficiency [1,3,5]. These hydrogen atoms in the physisorbed state, are weakly bound and immediately form stronger Pd-H bonds by hybridizing with Pd electronic states (chemisorption). The binding energy for chemisorption is about 0.4-0.6 eV per $H_2$ [3,5,45]. These chemisorbed H atoms remain highly mobile at the Pd surface and subsequently migrate into the metallic lattice. Once inside the Pd-lattice, H atom preferentially occupies octahedral (O) interstitial sites over tetrahedral (T) sites, as the former has larger free volume and lower energy relative to the latter [3,8]. The T(O)- sites in an FCC unit cell are indicated in Fig. 2(b). Inside the metal lattice, H atoms diffuse from the region of high concentration to the region of low concentration following Fick's law of diffusion [46]. A concentration gradient is readily formed along the thickness direction upon $H_2$ exposure, allowing H migration into the bulk towards the substrate. In contrast, lateral diffusion is restricted as in-plane H concentration remains uniform due to the full exposure of the film surface to $H_2$ gas. In the FCC-Pd, H atoms normally diffuse through local hopping between the neighbouring O-sites via the metastable T-site as indicated schematically in Fig. 2(c). The ground state energy level for the T-site is comparable with the 3$^{rd}$ energy level of the neighbouring O site, thus allowing resonant tunneling even at low temperatures [8].

Figures 2(d) and (e) show the hydrogen-induced magnetic hysteresis changes in two representative $Co_xPd_{100-x}$ (x = at.% Co) alloy films: one is Pd-rich $Co_{28}Pd_{72}$(7 nm) and the other is Co-rich $Co_{41}Pd_{59}$(7 nm). The present EHE measurements were performed under a fixed 4% $H_2/N_2$ atmosphere to compare the hydrogen-induced hysteresis evolution of CoPd alloys and Co/Pd multilayers under identical and near-saturated exposure conditions. Our previous concentration-dependent studies revealed that the response evolves systematically with $H_2$ concentration between 0.2% and 4% in $H_2/N_2$ gas. For a representative $Co_{32}Pd_{68}$ alloy film, the normalized EHE response at 4 % $H_2/N_2$ was about 16% in the magnetization-saturated state (B=0.5 T), ~22% at remanence, and ~182% at a fixed field of −13 mT, with sensitivity higher than 500%/$10^4$ ppm below 0.2% $H_2$ at B = −13 mT [11]. We also addressed the response/recovery behaviour under cyclic $H_2$/air exposure, which revealed that recovery depends on the magnetic-field condition. At saturated fields, the signal fully recovers upon hydrogen release, whereas within the hysteresis loop, complete recovery may require re-magnetization [11]. Our kinetic studies further indicated that the hydride formation time ($t_{H,50}$) also showed composition dependence, varying from about 60 s in Pd-rich $Co_{10}Pd_{90}$ to a few seconds in $Co_{40}Pd_{60}$, while the lattice response time ($t_{T,50}$) was significantly slower (~ $10^5$ s) in Co-rich compositions [47]. Therefore, the present work focuses on the composition- and structure-dependent crossover of the magnetic response, while systematic concentration-dependent studies across the full alloy and multilayer range remain important for future device optimization. Part of the composition and thickness dependence of the EHE in CoPd alloys has been reported in our earlier work [11]; however, the present results are obtained from independently prepared samples grown on crystalline GaAs substrates and are analyzed here with a focus on hydrogen-induced magnetic evolution probed via EHE. Both samples show different EHE hysteresis responses in hydrogen; the former shows contraction of the EHE loop, marked by the decrease in the coercivity and remanence, whereas the latter shows expansion of the EHE loop with a corresponding increase in the loop parameters $H_c$ and $\rho_{h,rem}$. The EHE loop contraction in the Pd-rich $Co_{28}Pd_{72}$(7 nm) film can be attributed to the combined effects of hydrogen-induced electronic modifications and magnetoelastic strain effects.

In Pd-rich alloys, overall magnetism is due to the spin polarization of the Pd-4d-band induced by the exchange interaction with the neighboring magnetic Co atoms [1,48]. Because Pd has a strong spin-orbit coupling, the hybridization between the Co-3d and Pd-4d orbitals modifies the d-electron cloud at Co-sites. This hybridization results in an anisotropic orbital moment in which out-of-plane orbital moments become stronger than the in-plane component, *i.e.,* orbitals with out-of-plane character ($d_{z^2}$, $d_{xz}$, $d_{yz}$) become lower in energy (more populated), whereas orbitals with in-plane character ($d_{xy}$, $d_{x^2-y^2}$) become less favorable [24]. This out-of-plane orbital moment anisotropy in the presence of strong SOC in Pd yields the net spin alignment in the out-of-plane direction, thus producing a robust PMA even when the film is grown on an amorphous substrate that imposes negligible coherent strain [23,24]. Upon hydrogen absorption, the 1s orbital of H causes filling of the Pd 4d-band (minority band), which in turn suppresses the Pd moment and weakens the Co-Pd hybridization and SOC that generates PMA [1].

In addition to its electronic influence, hydrogen occupying interstitial sites produces a substantial lattice expansion (3.4–3.6 %) [1]. For thin films clamped to a rigid substrate, this expansion becomes highly anisotropic because lateral (in-plane) deformation is suppressed, causing the lattice to expand predominantly in the out-of-plane direction [3]. Under such substrate constraint, the hydrogen-induced lattice expansion is expected to generate in-plane compressive stress; however, in the present study, this effect is discussed qualitatively as no direct measurement of hydrogen concentration or lattice parameter change is available. The magnetoelastic energy term defined by $E_{MEC} = \frac{3}{2}\lambda_{111}\sigma sin^2\theta$, where $\lambda_{111}$ is the magnetostriction coefficient and σ is the stress along in-plane direction and θ is the angle between magnetization and stress axis, modulates perpendicular anisotropy depending on the sign of $\lambda_{111}$ [49]. In the Pd-rich CoPd alloys, where the (111)-oriented magnetostriction coefficient $\lambda_{111}$ is negative [27], the hydrogen-induced in-plane compressive stress favors in-plane magnetization and thus leads to a contraction of the EHE hysteresis loop. Both the hydrogen-induced electronic suppression of Pd-magnetic moment, arising from the filling of Pd 4d states, and the magnetoelastic response under in-plane compressive stress act in the same direction and favor a net contraction of EHE hysteresis loops in Pd-rich $Co_{28}Pd_{72}$ alloy film.

In contrast, for the Co-rich $Co_{41}Pd_{59}$ alloy film (Fig. 1(e)), the hydrogen-induced response is opposite, characterized by an expansion of the EHE hysteresis loop with the increase in $H_c$, $\rho_{h,rem}$, and S = $\rho_{h,rem}/\rho_{h,sat.}$ Unlike Pd, in the case of Co-rich alloys, the Pd-induced magnetic moment is smaller, and the overall magnetism of the alloy mainly arises from a dominant Co-Co exchange coupling rather than Pd-polarization effects. Therefore, hydrogen absorption-induced electronic suppression of Pd magnetic moment becomes comparatively smaller relative to the magnetoelastic contribution to hysteresis change. The overall hysteresis change can be described by the hydrogen-induced magnetoelastic effect. Unlike Pd-rich CoPd alloys, the Co-rich alloys have positive $\lambda_{111}$ [27,28], therefore, hydrogen-induced in-plane compressive stress causes the magnetization to align in the out-of-plane direction relative to stress direction (in-plane) to minimize the magnetoelastic anisotropy energy. This results in the hydrogen-induced PMA in the Co-rich compositions.

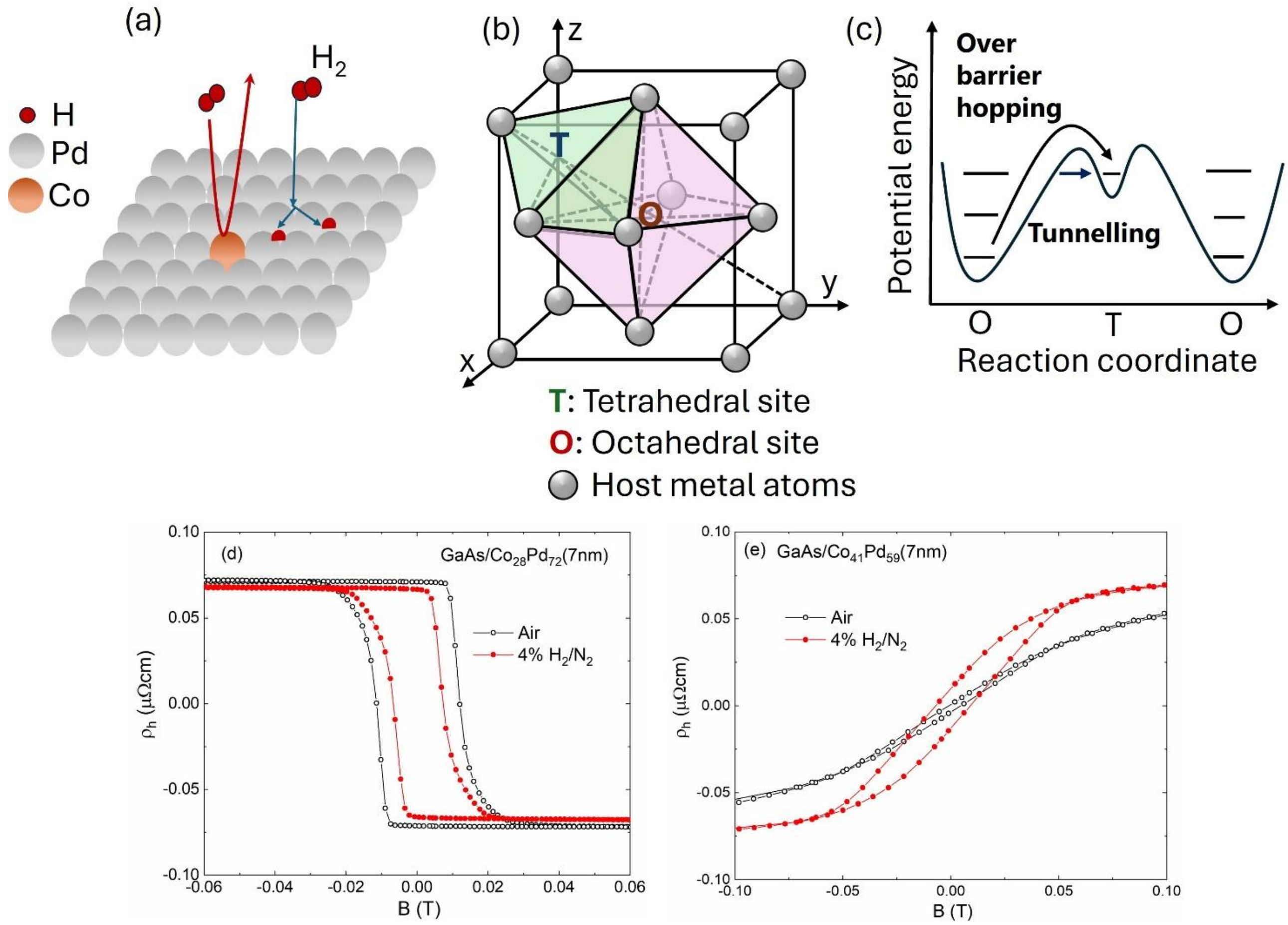

**Fig. 2** Hydrogen-induced structural and magnetic responses in CoPd alloys. (a) Schematics of dissociative adsorption of $H_2$ on CoPd surface and subsequent diffusion of atomic H into the bulk of the film. Red, grey, and orange spheres represent H, Pd, and Co atoms, respectively. (b) Tetrahedral (T) and Octahedral(O) interstitial sites in an FCC unit cell of CoPd alloys. The green and pink shaded regions denote the T and O sites, respectively, and grey spheres represent host metal atoms. (c) Potential energy diagram depicting the H diffusion through local hopping between neighboring O-sites via the metastable T site in CoPd sample. (d) Pd-rich $Co_{28}Pd_{72}$ alloy film showing EHE loop contraction under $H_2$. (e) Co-rich $Co_{41}Pd_{59}$ alloy film showing EHE loop expansion under $H_2$.

**(C) Composition-dependent hydrogen response in $Co_xPd_{100-x}$ alloys**

Figure 3(a) summarizes the composition dependence of the hydrogen-induced change in coercive field, $\Delta B_c = B_c(\mathrm{H_2}) - B_c(\mathrm{air})$, for 7 nm $Co_xPd_{100-x}$ alloy films grown on glass and crystalline GaAs substrates. The corresponding $B_c$ vs x plots in air and $H_2$ are shown in Fig. S2(a) and (b) in the supplementary file. A strong negative $\Delta B_c$ is observed for Pd-rich compositions (x ≤ 41 %), signifying pronounced hydrogen-induced magnetic softening and contraction of the EHE hysteresis loops. The magnitude of this effect is substantial, reaching values as large as $\Delta B_c = -15$ mT for x ~ 28 % (at.% Co), corresponding to a ~ 56% reduction relative to the hydrogen-free state. The corresponding relative change $\Delta B_c/B_c$ is plotted as a function of x in Fig. S2(c) in the supplementary. This underscores the high sensitivity of Pd-dominated magnetism to hydrogen absorption in Pd-rich compositions.

With increasing Co concentration, $\Delta B_c$ steadily increases and undergoes a clear sign reversal between x = 30 % and 40 at.%. In the Co-rich regime x ≥ 40 %, the hydrogen response is strongly reduced, and $\Delta B_c$ becomes weakly positive or nearly zero, indicating a transition to hydrogen-induced loop expansion or a largely hydrogen-insensitive magnetic state. The crossover from loop contraction to loop expansion in CoPd is gradual rather than abrupt. This is inferred from the continuous evolution of $\Delta B_c$ and ΔS with increasing Co concentration, from large negative values in the Pd-rich regime to weakly positive or nearly zero values near and above x ~ 40 %. Thus, the crossover represents a relative balance between Pd-derived electronic suppression and strain-mediated magnetoelastic effects. The Pd-derived electronic suppression is consistent with the DFT studies on Pd–H systems, which suggest that hydrogen absorption induces strong Pd 4d–H 1s bonding, modifies Pd d-state filling, and reduces the density of states near the Fermi level [36]. Concentration-dependent EHE loop measurements in CoPd films were reported in our earlier works [11, 14], which show gradual evolution of EHE loops with increasing $H_2$ concentration. A representative $H_2$-concentration-dependent EHE response for Glass/$Co_{36}Pd_{64}$(7 nm) is provided in Fig. S5, further showing the systematic evolution of the EHE response with $H_2$ concentration. Therefore, in the present work, 4% $H_2/N_2$ was used as a fixed exposure condition to compare the composition-dependent hydrogen response. The initial magnetic anisotropy state should also be considered in understanding the composition-dependent hydrogen response. Pd-rich films exhibit relatively square hysteresis loops, indicating stronger PMA, whereas Co-rich films show more tilted loops, suggesting a larger in-plane anisotropy (IPA) component. Previous studies have highlighted that hydrogen can influence magnetic anisotropy, spin-reorientation behavior, and magnetic domain structures in CoPd alloys and Co/Pd multilayers [12,16,50]. Thus, although the hydrogen-induced hysteresis evolution in CoPd is mainly composition dependent, the detailed loop response can also depend on the initial magnetic state, as hydrogen modifies an already existing balance among PMA, IPA, domain reversal and pinning, in addition to causing electronic and magnetoelastic modifications. The crossover composition (x ~ 38 %), highlighted by the vertical dashed line, closely coincides with the reported sign change of the (111) magnetostriction coefficient $\lambda_{111}$ in CoPd alloys [27,28], suggesting a fundamental link between magnetostriction and the hydrogen response. The films grown on GaAs substrates show qualitatively similar behavior with only minor quantitative differences due to differences in strain, microstructure, and interface quality, consistent with Ref. [11]. Hydrogen absorption can also be influenced by the film thickness, grain boundaries, defect density, and interface structure. Although these factors can introduce quantitative variations among samples, the systematic sign change of $\Delta B_c$ and $\Delta S$ with composition observed in films with identical thickness and growth conditions on different substrates (glass/GaAs), indicates a primarily composition-driven crossover. Although direct experimental separation of electronic and magnetoelastic contributions was not performed here, future in situ XPS, XMCD, XRD, or strain-sensitive measurements under hydrogen exposure would be interesting for directly correlating electronic modification, strain evolution, and magnetic anisotropy in CoPd alloys and Co/Pd multilayers. In contrast to the 7 nm films, $\Delta B_c$ for the thicker (14 nm) $Co_xPd_{100-x}$ films remains mostly negative across the composition range, with no evidence of sign reversal. This indicates that the hydrogen-induced magnetic response becomes more uniform at larger thickness, where Pd-driven electronic effects dominate over the magnetoelastic effects.

Additionally, we also extracted the squareness (S) of the EHE loops defined by $S = \rho_{h,\mathrm{rem}}/\rho_{h,\mathrm{sat}}$ and its changes $\Delta S = S(\mathrm{H_2}) - S(\mathrm{air})$, in hydrogen. Figure 3(b) shows the hydrogen-induced change in ΔS for various x values. In the Pd-rich alloys, hydrogen exposure leads to a pronounced reduction in squareness ($\Delta S < 0$), consistent with a weakening of perpendicular magnetic anisotropy and a more gradual, reversible magnetization reversal process. As the Co concentration increases, the magnitude of $\Delta S$ diminishes and approaches zero in the Co-rich regime (x > 45 %), indicating that the remanent magnetization and loop shape become increasingly insensitive to hydrogen. The similar trends observed in $\Delta B_c$ and $\Delta S$ demonstrate that hydrogen-induced modifications in both the coercive field and PMA in the CoPd system are composition dependent. Here, the loop contraction and loop expansion are defined in terms of experimentally observed hydrogen-induced changes in the hysteresis parameters. Loop contraction corresponds to a negative $\Delta B_c$ and ΔS, whereas loop expansion refers to a positive $\Delta B_c$ and ΔS. Microscopically, depending on composition and microstructure, the loop expansion under hydrogen may reflect enhanced PMA, broader switching-field distribution, or increased domain-wall pinning. Together, these results establish a clear crossover from a Pd-dominated regime, where hydrogen strongly suppresses magnetism, to a Co-dominated regime, where the hydrogen-induced magnetic response is markedly weaker or reversed.

While the alloy series reveals a clear composition-controlled crossover in hysteresis evolution, Co/Pd multilayers provide a complementary platform to explore hydrogen-induced magnetic effects under the influence of interfacial hybridization, reduced dimensionality, and magnetic connectivity.

Moreover, the interplay between the hydrogen-induced electronic and magnetoelastic effects in the overall magnetic state of the system is of significant interest for both fundamental physics and device applications.

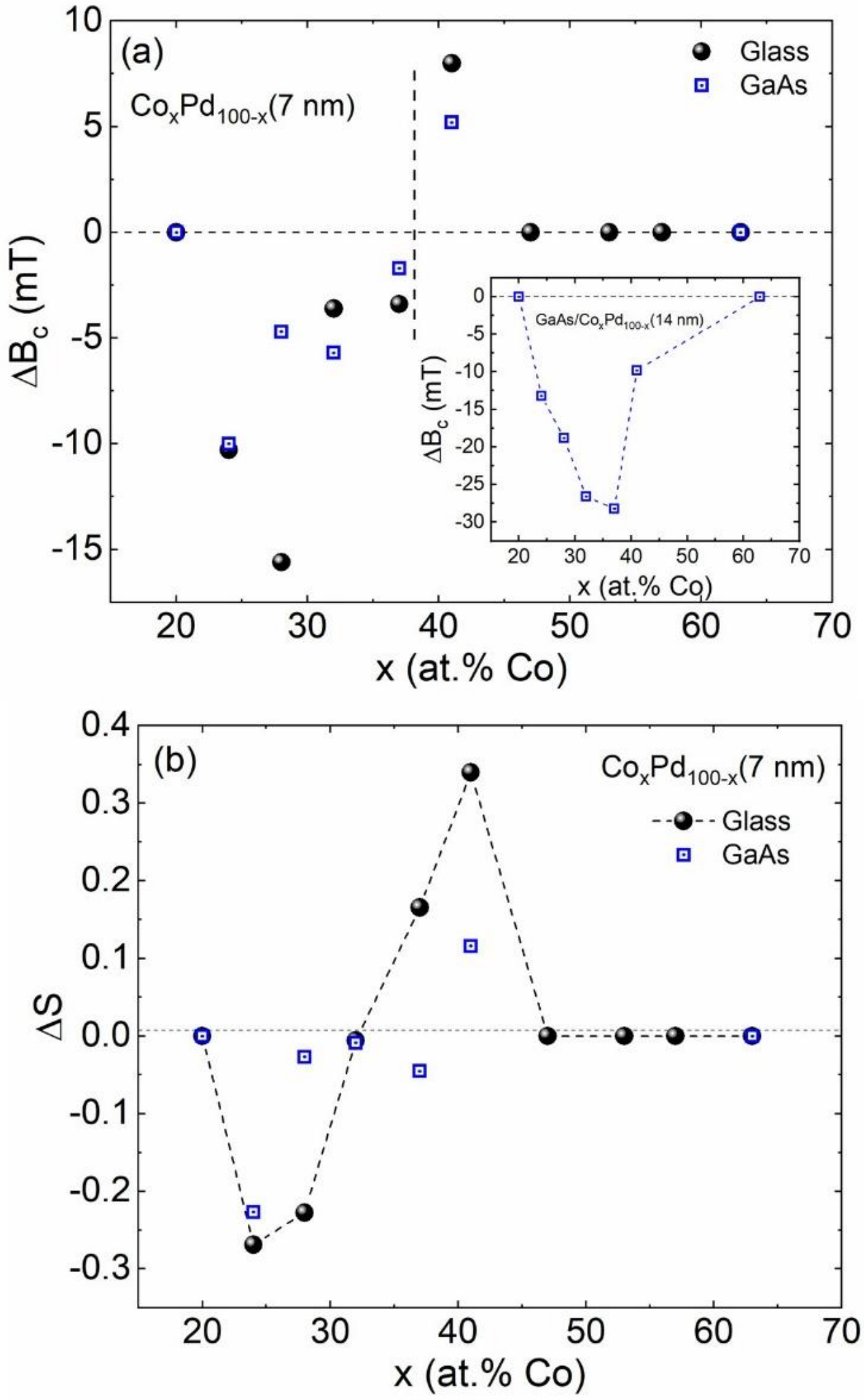

**Fig. 3** Composition-dependent hydrogen-induced hysteresis evolution in $Co_xPd_{100-x}$ alloy films. (a) Change in coercive field, $\Delta B_c = B_c(H_2) - B_c(air)$, as a function of Co concentration x for 7 nm films grown on glass and GaAs substrates. Pd-rich compositions exhibit negative $\Delta B_c$, indicating hydrogen-induced loop contraction, whereas Co-rich compositions show a suppressed or weakly positive response. The vertical dashed line marks the approximate crossover composition near x ~ 40 %. The inset shows

$\Delta B_c(x)$ for a 14 nm film, highlighting the disappearance of the sign reversal with increasing thickness. (b) Corresponding change in loop squareness, $\Delta S = S(H_2) - S(air)$ , where $S = \rho_{h,rem}/\rho_{h,sat}$ . Negative $\Delta S$ in Pd-rich alloys reflects hydrogen-induced weakening of PMA, while Co-rich compositions show minimal change.

**(D) Hydrogen-induced hysteresis evolution in Co/Pd multilayers**

Figure 4 compares the hydrogen-induced hysteresis changes in two ultrathin multilayer series, $[Co(0.1\ nm)/Pd(d)]_{15}$ and $[Co(0.2\ nm)/Pd(d)]_{15}$, in which the effective Co volume fraction $x$ is tuned by varying the Pd layer thickness $d$. The number of bilayers in both series is fixed at 15. When compared to the alloy series, these multilayers exhibit PMA driven by an interfacial Co-Pd hybridization and spin-orbit coupling in the presence of broken symmetry at the interface, consistent with X-ray magnetic circular dichroism (XMCD) and first-principles studies on Co/Pd multilayers [26]. In addition, partial intermixing at the Co/Pd interfaces can lead to effective alloy-like behavior, as commonly reported in Co/Pd multilayer systems [51]. Therefore, the multilayers are discussed in terms of effective Co fraction, interface effects, and magnetic connectivity, rather than as continuous Co/Pd layers. In these multilayers, interface quality, intermixing, and Co-layer thickness can also affect hydrogen uptake and strain distribution. Although such structural factors can influence the magnitude of hydrogen response, the observed systematic trend mainly reflects the composition-dependent balance between Pd-derived electronic and magnetoelastic contributions. The extent of this intermixing typically depends on growth conditions. The magnetic coupling strength can be systematically enhanced by increasing the Co-layer thickness. Across both series, the EHE loops exhibit a clear hydrogen-induced hysteresis evolution, progressing from a weak hysteresis response at very low effective Co fraction (x) to a more pronounced hysteresis change with either loop contraction or expansion, depending on x and Co thickness.

In the [Co(0.1 nm)/Pd(d)]$_{15}$ multilayers (panels a–c, left column), the lowest effective Co fraction (x ~ 10 %) shows a nonlinear Hall response with negligible hysteresis, and hydrogen produces only a minor change of the loop (Fig. 4(a)). This trend is consistent with a dilute regime in which the Co sublayers are discontinuous, such that long-range ferromagnetic order and stable magnetic domains are not fully developed. Upon increasing the effective Co fraction to x ~ 22 %, clear perpendicular hysteresis appears, and hydrogen exposure leads to a distinct expansion of the EHE loop (Fig. 4(b)), marked by the corresponding increase in the loop width. This expansion can be ascribed to the hydrogen-induced magnetoelastic effect, where hydrogen absorption in Pd layers generates in-plane compressive stress that enhances the effective PMA in the sample. At the same time, $\rho_{h,rem}$ shows a slight decrease, indicating that hydrogen modifies the magnetization reversal and domain configuration apart from enhancing the PMA. With a further increase of the effective Co fraction to $x \approx 40$ % (Fig. 4(c)), the EHE hysteresis, which is originally absent in air, becomes pronounced in hydrogen, indicating further hydrogen-induced enhancement of PMA. Furthermore, as evident from Figs. 4(a–c), the effective Hall amplitude decreases over the entire field range, consistent with hydrogen-induced suppression of the effective EHE amplitude, arising from reduced Pd polarization and/or a reduced EHE coefficient. This mixed response indicates that, near the crossover composition, hydrogen activates both magnetoelastic (anisotropy-related) and electronic (band-filling related) channels, which affects the EHE loop width and the Hall amplitude, respectively. This interpretation is consistent with previous first-principles and in operando PNR/FMR studies on Co/Pd systems, which showed that hydrogen can modify local Co–Pd hybridization, weaken the PMA associated with the Co/Pd interface, induce Pd-layer expansion, and alter spin polarization near the Fermi level [13,32].

In contrast, the [Co(0.2 nm)/Pd(d)]$_{15}$ multilayers (Fig. 4(d)-(f)) show different behaviour, highlighting the role of magnetic connectivity and Co-dominated exchange. For x ~ 10 % (Fig. 4(d)), the loops are weakly hysteretic and largely field-reversible, indicating that the effective Co fraction remains too small to produce robust PMA. However, upon increasing x to 31 % (Fig. 4(e)), a well-defined perpendicular hysteresis appears, and hydrogen modifies the EHE loop with significant reduction in $B_c$ and $\rho_{h,rem}$. Importantly, at a higher effective Co fraction (x ~ 60%, Fig. 4(f)), hydrogen produces an opposite trend compared to the Pd-rich regime (low x) within the same multilayer series: the EHE response shows a larger Hall amplitude and a more pronounced loop opening around zero field. This suggests that hydrogen increases both the amplitude of $\rho_H$ and strengthens PMA in this Co-rich multilayer. Such behaviour is consistent with a crossover toward Co-dominated magnetism where Co-Co exchange and magnetic connectivity are stronger, while Pd plays a minor role in setting the net magnetism. In this regime, hydrogen-induced magnetoelastic contribution becomes dominant, which, in the presence of an in-plane compressive stress, induces perpendicular spin alignment in Co-rich multilayers with positive $\lambda_{111}$. Simultaneously, hydrogen may modify the EHE coefficient $R_s$ through band-filling and scattering changes, which can further amplify the EHE amplitude. Overall, the multilayer comparison indicates a thickness-controlled crossover: increasing the Co sublayer thickness from 0.1 to 0.2 nm shifts the hydrogen response from predominantly loop expansion to loop contraction in Pd-rich regime ($x < 40$ %). This multilayer behavior complements the alloy series by showing that the same electronic and magnetoelastic contribution in CoPd alloys also operates in multilayer nanostructures, but their relative weight is strongly modulated by interfacial hybridization and magnetic connectivity at the interface.

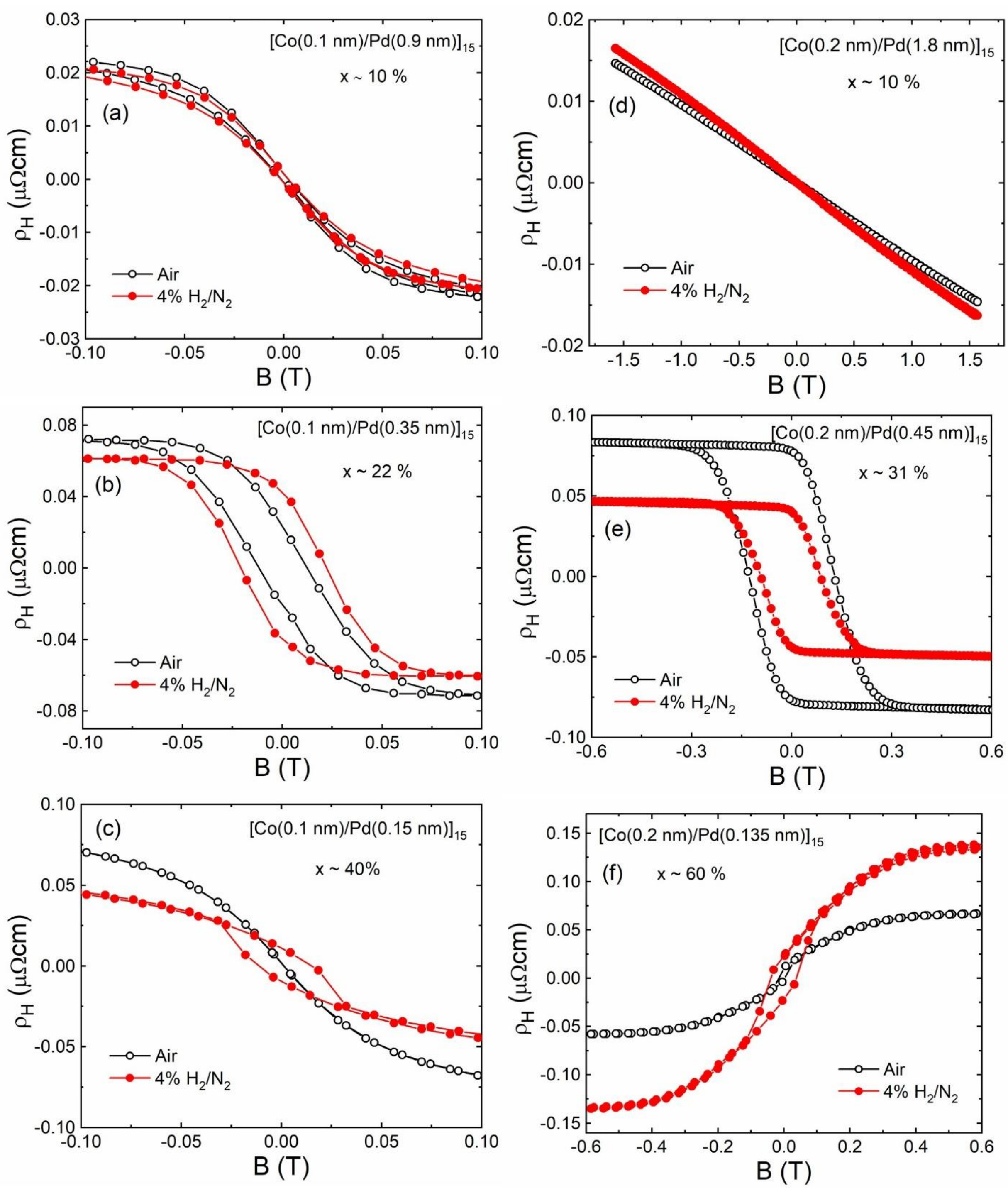

**Fig. 4** Hydrogen-induced evolution of EHE hysteresis loops in Co/Pd multilayers. Representative EHE-loops measured in air and under 4% $H_2/N_2$ mixture for $[Co(0.1\ nm)/Pd(d)]_{15}$ multilayers with effective Co volume fractions (a) x ~ 10 %, (b) 22 %, and (c) 40 %, and for $[Co(0.2\ nm)/Pd(d)]_{15}$ multilayers with (d) x ~ 10 %, (e) 31 %, and (f) 60 %. These figures highlight the thickness(composition)-dependent

interplay between hydrogen-induced magnetoelastic anisotropy and electronic effects in the evolution of EHE hysteresis loops.

**(E) Quantitative hydrogen response in Co/Pd multilayers**

Figure 5 summarizes the hydrogen-induced changes in the hysteresis parameters in Co/Pd multilayers with different Co thicknesses. Fig. 5(a) shows the variation of $B_c$ as a function of Co volume fraction for the $[Co(0.2\ nm)/Pd(d)]_{15}$ multilayers. A decrease in the $B_c$ values in $H_2$ is observed below x ~ 45 %, indicating hydrogen-induced magnetic softening. In contrast, at higher Co fractions (x > 45 %), $B_c$ increases under hydrogen exposure, reflecting a transition to hydrogen-induced magnetic hardening. Corresponding $B_c$ vs x plots for $[Co(0.1\ nm)/Pd(d)]_{15}$ multilayer is shown in Fig. S3(a). The normalized $B_c$ ($\Delta B_c/B_c$) data are also provided in the Supplementary (Fig. S3(b)) for completeness. Figures 5(b) and 5(c) show the hydrogen-induced changes in coercive field, $\Delta B_c$, and squareness, $\Delta S$, respectively, as a function of effective Co volume fraction for the $[Co(0.1\ nm)/Pd(d)]_{15}$ and $[Co(0.2\ nm)/Pd(d)]_{15}$ multilayers. In the first Co(0.1 nm) series (Fig. 5(b)), $\Delta B_c$ exhibits a weak and non-monotonic hydrogen response, with both positive and small negative values depending on composition. At low Co-fractions (x ≤ 20 %), $\Delta B_c$ is nearly zero, indicating that hydrogen only slightly affects the magnetization reversal. Upon increasing x, $\Delta B_c$ becomes slightly positive (loop expansion) over a broad intermediate range (x ~ 20–45 %), and shows an increase up to x = 60 %, above which it becomes zero. Similar to $\Delta B_{c,}$, $\Delta S$ also remains positive and modest at low and intermediate x, indicating hydrogen causes a slight stabilization of domains in the perpendicular direction. Together, these results show that even in interface-dominated ultrathin Co(0.1nm) multilayers, hydrogen produces a net weak expansion of loops, arising as compensation between electronic and magnetoelastic effects.

In contrast, the thicker Co(0.2 nm) multilayers exhibit a relatively stronger and more pronounced hydrogen response. In low Pd-rich fractions (x ~ 20 -45 %), $\Delta B_c$ becomes negative, reaching a pronounced minimum around x ~ 35 %, accompanied by a negative $\Delta S$. This indicates hydrogen-induced magnetic softening and loop contraction in Pd-rich fractions produced as a consequence of electronic suppression of Pd-induced moment. With increasing x further ($x \geq 50$ %), both $\Delta B_c$ and $\Delta S$ change sign and become positive, indicating a crossover to hydrogen-induced loop expansion and enhanced PMA due to the dominance of the magnetoelastic effect. Notably, such a transition occurs at a relatively higher x than in Co(0.1 nm) series, indicating that thicker Co sublayers require a larger Co content for the magnetoelastic contribution to overcome the Pd-related electronic effects. This shift reflects the increased role of Co-Co exchange and magnetic connectivity in hydrogen-induced modification of the Pd-induced magnetic moment.

Overall, Figure 5 reveals a thickness-controlled crossover of hydrogen-induced hysteresis response in Co/Pd multilayers. While ultrathin Co-layers cause a weak expansion of the hysteresis loops, increasing the Co thickness shifts the regime towards higher x. These results suggest that hydrogen response in Co-Pd nanostructures is governed not only by composition, but by Co-sublayer thickness, magnetic connectivity, and interfacial hybridization, providing a powerful route for engineering hydrogen-responsive spintronic devices.

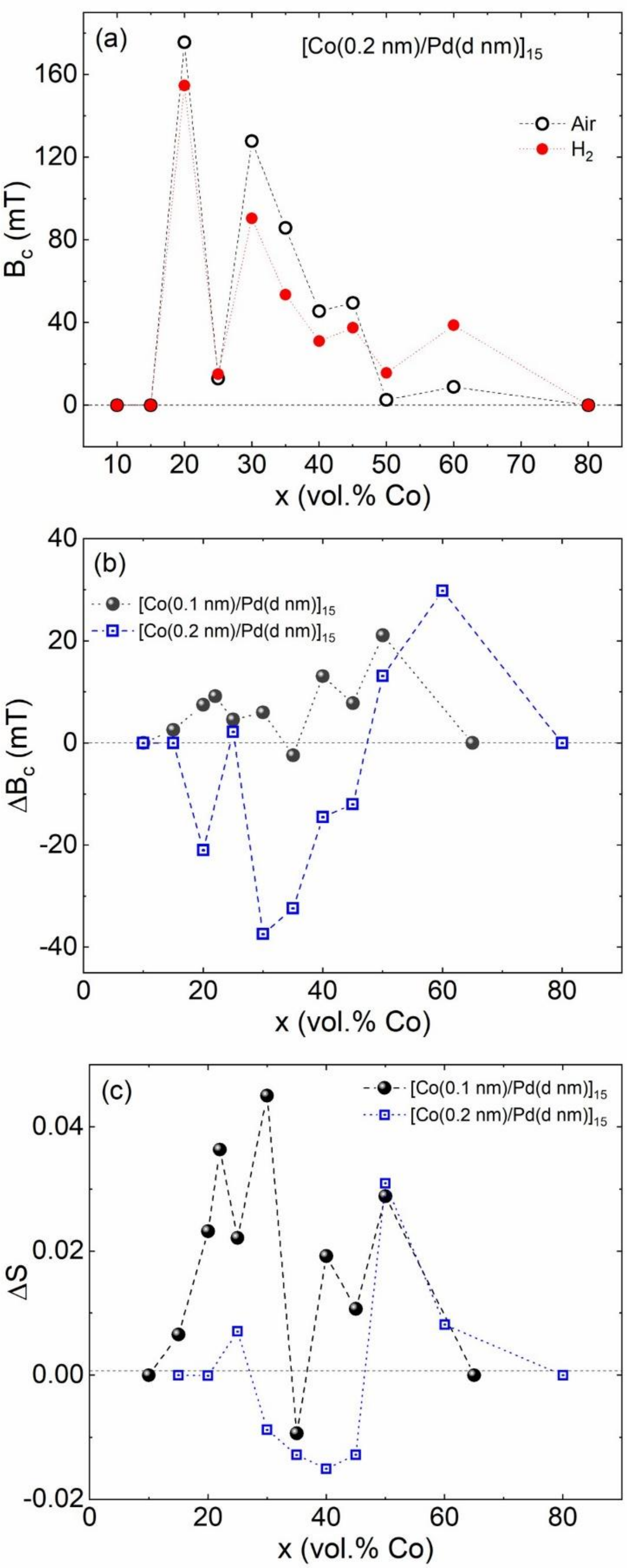

(a)
[Co(0.2 nm)/Pd(d nm)]15
Air
H2
Bc (mT)
x (vol.% Co)
(b)
[Co(0.1 nm)/Pd(d nm)]15
[Co(0.2 nm)/Pd(d nm)]15
ΔBc (mT)
x (vol.% Co)
(c)
[Co(0.1 nm)/Pd(d nm)]15
[Co(0.2 nm)/Pd(d nm)]15
ΔS
x (vol.% Co)

**Fig. 5** Composition-dependent hydrogen response in Co/Pd multilayers with different Co-thicknesses. (a) Coercivity $B_c$ as a function of x for the $[Co(0.2\ nm)/Pd(d)]_{15}$ multilayer in air and 4 % $H_2/N_2$. (b) $\Delta B_c$ as a function of x for $[Co(0.1\ nm)/Pd(d)]_{15}$ (black solid circles) and $[Co(0.2\ nm)/Pd(d)]_{15}$ (blue squares) multilayer series. (c) ΔS as a function of x for the two-multilayer series. Dashed and dotted lines serve as guides to the eye. Thicker Co layers exhibit predominantly negative $\Delta B_c$ for $x < 40$ % (Pd-rich), indicating hydrogen-induced loop softening, whereas ultrathin Co layers show positive $\Delta B_c$, corresponding to hydrogen-induced loop expansion. The trends reveal a thickness-controlled crossover from a Pd-dominated to a Co-dominated magnetic response.

**(F) Hydrogen-induced EHE polarity reversal**

In addition to modifying the hysteresis characteristics, hydrogen induces a reversal of the EHE loop polarity from negative to positive once the effective Co concentration exceeds approximately 35% in the CoPd systems. In the case of $Co_{37}Pd_{63}$ (7 nm) alloy film (Fig. 6a), as reported in our earlier work [11], the EHE signal measured in air exhibits a negative slope and negative high-field Hall resistivity, whereas upon exposure to 4% $H_2/N_2$, the loop polarity inverts, yielding a positive Hall signal over the same field range. This polarity reversal occurs without a drastic change in the overall loop shape, indicating that hydrogen primarily alters the sign of the anomalous Hall coefficient $R_s$ rather than the basic magnetization reversal mechanism.

Similarly, in the case of Co/Pd multilayers, a more pronounced EHE reversal is observed in Co-rich fractions. For the $[Co(0.1\ nm)/Pd(0.055\ nm)]_{15}$ multilayers stack with an effective Co fraction x ~ 65 % (Fig. 6(b)), hydrogen absorption yields a complete reversal of the EHE sign along with the enhancement of EHE amplitude. Such behaviour demonstrates a transition from Pd-rich to Co-dominated magnetic regime in which hydrogen effectively modifies the $R_s$. Hydrogen-induced band filling and lattice strain alter the balance between intrinsic (Berry-curvature-driven) and extrinsic (skew-scattering and side-jump) EHE contributions, leading to a sign change of $R_s$, and hence the EHE response. It is to be noted here that, at high Co content (x ≈ 65%), when the Pd layer thickness approaches sub-monolayer coverage (d ~ 0.055 nm), the partial intermixing at the interface could lead to alloy-like regions. However, due to sequential deposition, which can still create local Co-rich and Pd-rich regions, the multilayer is expected to retain local compositional modulation and interface effects, even with partial intermixing. This can result in differences in magnetic response compared to a homogeneous alloy.

Similar behavior is also observed in the $[Co(0.2\ nm)/Pd(0.2\ nm)]_{15}$ multilayer with x = 50 % (Fig. 6c), where the hydrogen incorporation produces pronounced EHE-polarity reversal. After hydrogen removal, the "air-last" loop does not reproduce the initial loop, indicating the recovery is only partial under the present desorption condition. This suggests that, in addition to a reversible electronic contribution, hydrogen also causes a persistent modification of magnetic/electronic state due to residual hydrogen, modified domain configuration, or slow relaxation, which changes the EHE amplitude.

Overall, these results show that hydrogen can not only modulate the EHE loop hysteresis but can also reverse its sign in CoPd systems near and beyond the Co-rich regime. The polarity reversal is attributed to hydrogen-induced modifications of the EHE scattering mechanisms(skew scattering/side-jump) and their relative weight, band filling, spin-orbit scattering, etc. This polarity reversal is highly sensitive to Co fraction in nanostructured alloys and multilayers. Additionally, the hydrogen-induced EHE polarity reversal provides an additional and previously unexplored degree of control offered by hydrogen interaction in Pd-based magnetic nanostructures.

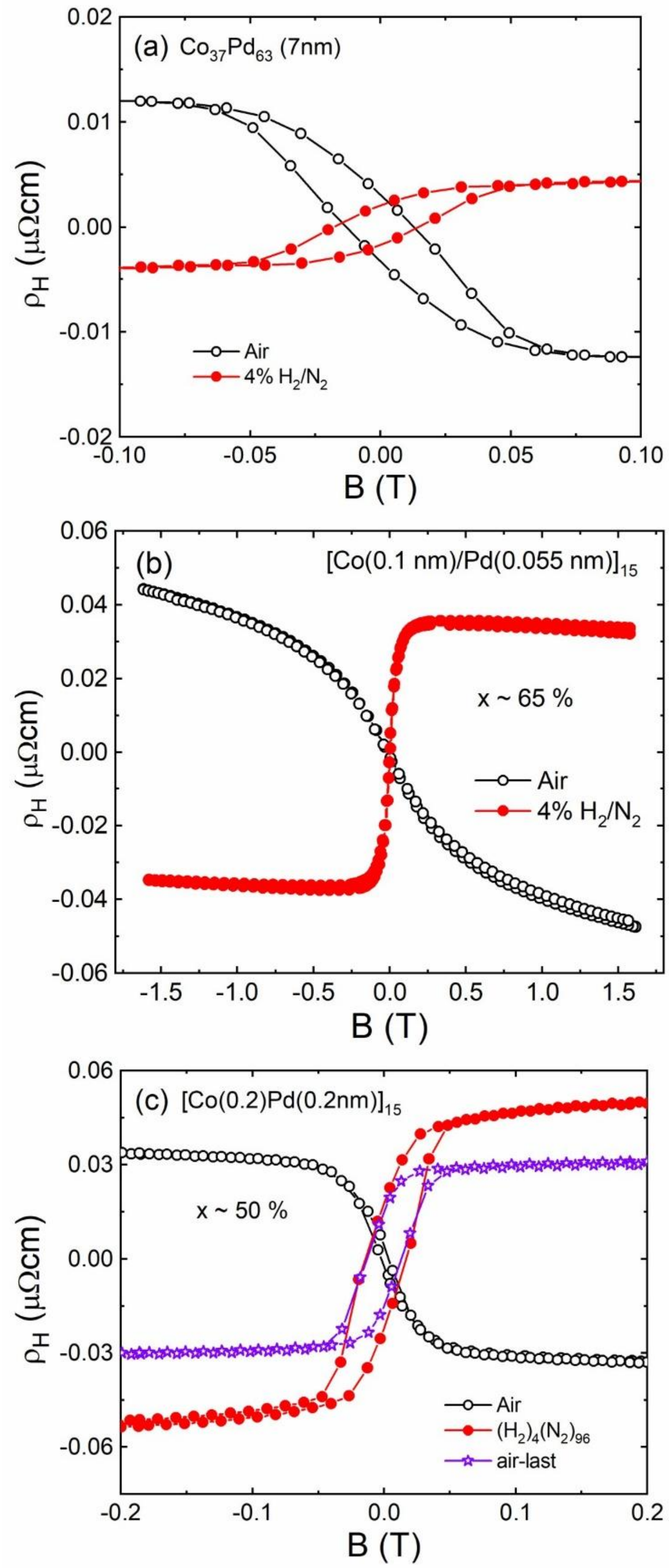

**Fig. 6** Hydrogen-induced EHE polarity reversal. Representative EHE hysteresis loops measured in air and under 4% $H_2/N_2$ for (a) a $Co_{37}Pd_{63}$ (7 nm) alloy film (adapted from Ref. [11]), (b) a Co-rich multilayer

$[Co(0.1\ nm)/Pd(0.055\ nm)]_{15}$ with effective Co fraction x ~ 65 % and (c) a $[Co(0.2\ nm)/Pd(0.2\ nm)]_{15}$ multilayer with x ~ 50 %. In these samples, hydrogen exposure induces a reversal of the EHE polarity while preserving hysteretic behavior, indicating a hydrogen-driven sign change of the EHE coefficient.

**(G) EHE scaling and hydrogen-induced sign reversal**

Hydrogen-induced modifications in the EHE resistivity are further analyzed based on scaling analysis of EHE resistivity $\rho_h$ with the longitudinal resistivity ρ for the $[Co(0.2\ nm)/Pd(d)]_{15}$ multilayers in Fig. 7. The corresponding plots for the $[Co(0.1\ nm)/Pd(d)]_{15}$ multilayers are provided in the Supplementary Material (Fig. S4). Figure 7(a) shows the magnitude of EHE resistivity $|\rho_h|$ as a function of ρ, measured in air and under 4 % $H_2$. Instead of following a simple monotonic scaling behaviour ($|\rho_h| \propto \rho^n$), the data exhibit a nontrivial and scattered distribution. In conventional EHE scaling, skew scattering usually gives a linear relation between $\rho_h$ and $\rho$, whereas intrinsic Berry curvature and side-jump dominance give a quadratic dependence as $\rho_h \propto \rho^2$ [52,53]. Therefore, a single power-law dependence can only be possible when one mechanism dominates over the entire composition range. The hydrogenated state shows a systematic shift of the data points relative to air, suggesting modification of $\rho_h$ cannot be explained solely by the changes in the scattering of carriers. Moreover, the absence of a pure power-law dependence suggests that $H_2$ induces multiple contributions to the EHE, which include intrinsic (Berry curvature-related) and extrinsic mechanisms such as skew scattering and side-jump processes, along with modifications arising from changes in magnetization, magnetization reversal and magnetoelastic effects [16,52,21]. Hydrogen-induced modification of electronic structure and strain effect can therefore change the relative weight of these intrinsic and extrinsic EHE contributions, leading to the observed nonmonotonic scaling and polarity reversal.

In Fig. 7(b), the $H_2$-induced changes in the Hall resistivity $\Delta\rho_h$ and longitudinal resistivity ($\Delta\rho = \rho(H_2) - \rho(air)$) are further correlated. The data exhibit nonlinear and widely dispersed distributions, indicating that conventional scaling relations are not applicable. Since EHE measurements in multilayers can be influenced by current shunting, the EHE response was interpreted together with the longitudinal resistivity change. The absence of a simple linear or power-law relationship between $\Delta\rho_h$ and $\Delta\rho$ indicates that the EHE evolution cannot be explained solely by hydrogen-induced resistivity changes or current redistribution; rather, it reflects the interplay of multiple hydrogen-modified contributions. In addition to intrinsic and extrinsic EHE mechanisms, hydrogen absorption can also influence the magnetic domain evolution and domain-wall motion in CoPd system, which in turn can contribute to the observed nonmonotonic behavior. Kerr microscopy-based studies on CoPd alloys and Co/Pd multilayers have shown that hydrogen can modify magnetic domain formation, anisotropy, and spin-reorientation behavior [12,13,17]. Since $\rho_h$ ($\sim R_s 4\pi M$) also depends on the perpendicular component of magnetization, any change in domain nucleation, domain-wall pinning, and reversal pathways can alter the remanence, coercivity, and apparent Hall amplitude. Therefore, the nonlinear $\Delta\rho_h$–$\Delta\rho$ relationship likely reflects combined modifications of carrier scattering, magnetic anisotropy, magnetization reversal, and domain-wall behavior under hydrogen exposure.

Figure 7(c) shows the $\rho_h$ as a function of x in air and $H_2$. In air, $\rho_h$ mostly remains negative up to x ~ 55 %, above which it changes to positive polarity. In contrast, under $H_2$, $\rho_h$ shows a positive shift and undergoes sign reversal at a lower x (~ 45 %) compared to the air value. The crossing of $\rho_h = 0$ indicates that hydrogen changes the dominant mechanisms governing the EHE, reflecting simultaneous modifications in the electronic structure, magnetoelastic contributions, and scattering processes. Such a transition signifies a direct link between composition, hydrogen incorporation, and the electronic structure of the system. Moreover, the interplay of multiple hydrogen-induced contributions leads to the observed non-monotonic scaling behavior of EHE.

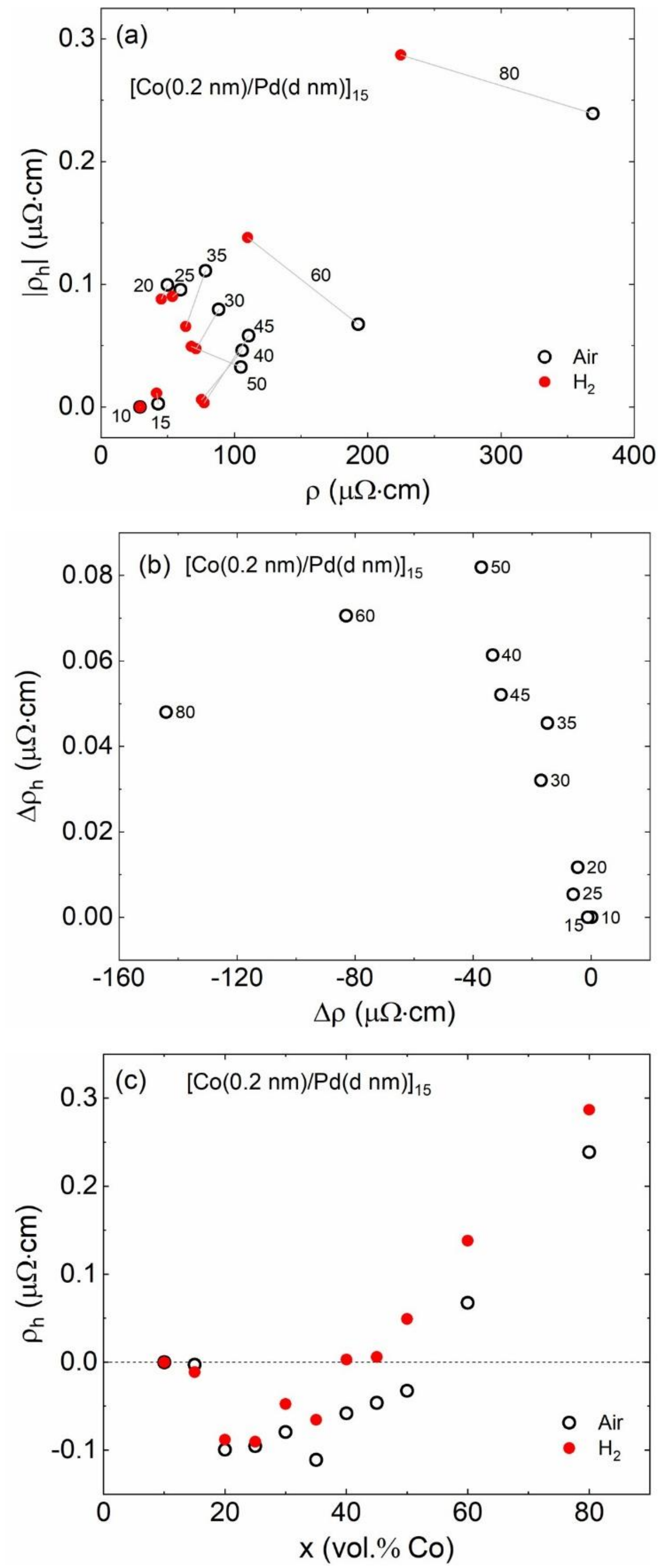

**Fig. 7** (a) Scaling of the EHE resistivity, $|\rho_h|$, as a function of longitudinal resistivity, $\rho$, for [Co(0.2 nm)/Pd(d)]$_{15}$ multilayers measured in air and 4% $H_2/N_2$. Here, the lines are guides to the eyes, connecting corresponding air and $H_2$ data points for each Co concentration. (b) Hydrogen-induced changes in Hall resistivity ($\Delta\rho_h$) as a function of longitudinal resistivity ($\Delta\rho$), showing a clear deviation from linear scaling. (c) EHE resistivity $\rho_h$ as a function of Co concentration x for $[Co(0.2\ nm)/Pd(d\ nm)]_{15}$ multilayers. The horizontal dashed line shows $\rho_h = 0$. A hydrogen-induced sign reversal of the EHE is observed at the intermediate Co-concentrations (40–50 %).

**(H) Hydrogen response regimes in Co-Pd nanostructures**

The overall hydrogen-induced response regimes in CoPd systems can be represented in terms of different response regimes (Fig. 8), owing to the competition between electronic and magnetoelastic effects. These regimes are predominantly dependent on the effective Co fraction, microstructure, and thickness of the sublayers, nature of the interface, etc. In the Pd-rich regime, the hydrogen-induced response is mainly dominated by the suppression of the electronic effect due to band filling of Pd-4d band, which results in the hysteresis loop contraction. As the effective Co fraction increases or the Co sublayers become thicker and magnetically connected, the system transitions to a crossover regime in which both electronic suppression and magnetoelastic contributions coexist, resulting in mixed behaviour in EHE hysteresis response. At higher Co fractions ($x > 40$ %), a Co-dominant regime emerges in which hydrogen affects magnetic anisotropy via magnetoelastic coupling. In this regime, hydrogen absorption induces PMA in the samples, resulting in the expansion of EHE loops. The electronic effects play a minor role. Notably, the crossover between these regimes shifts systematically by Co thickness in multilayers, demonstrating that hydrogen-induced magnetic response is strongly dependent on the nanoscale structure, composition, and interface hybridization. Further, the crossover concentration for alloys (~ 38 %) is lower than that for multilayers (~ 46 %), reflecting the enhanced role of interface and strain effects, which extend the Pd-dominated electronic response to higher Co fractions. Future micromagnetic simulation or first-principles modeling would be helpful to quantitatively distinguish the electronic, magnetoelastic, and scattering contributions responsible for the hysteresis transition and EHE polarity reversal in the CoPd system. These results provide a unified basis for engineering hydrogen-responsive magnetic behavior in CoPd nanostructures.

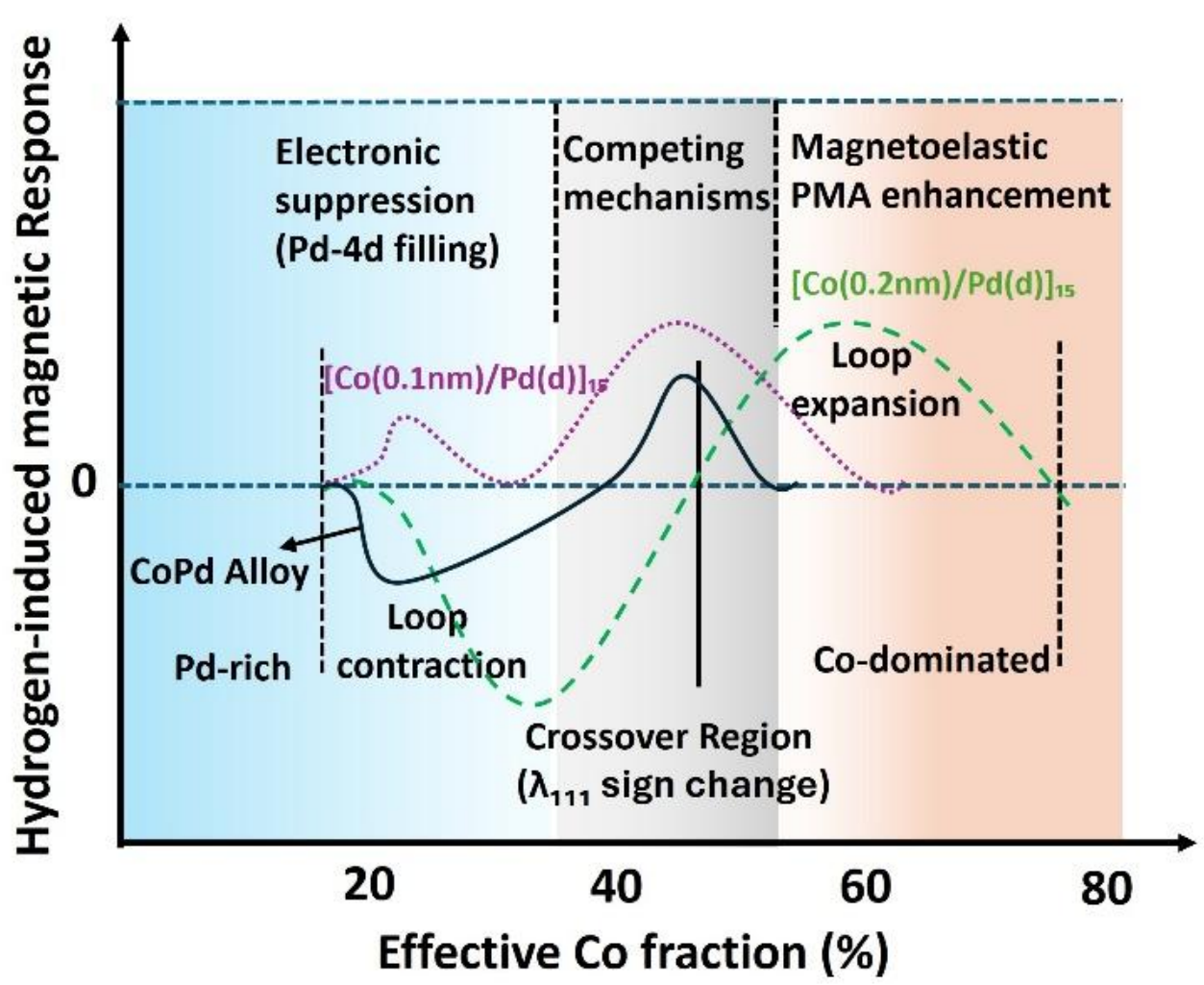

**Fig. 8** Schematic of hydrogen-induced magnetic response as a function of effective Co-fraction in CoPd systems. The solid line represents the baseline behaviour of CoPd alloy films, while the dashed curves correspond to multilayer systems with different Co thicknesses. The diagram illustrates the crossover from a Pd-rich regime dominated by hydrogen-induced electronic suppression and hysteresis contraction to a Co-dominated regime where magnetoelastic enhancement of PMA leads to loop expansion. The crossover region reflects competition between these mechanisms and is tuned by the effective Co fraction and magnetic connectivity. Multilayers exhibit enhanced effects compared to alloys due to interface and thickness effects.

## IV. Conclusion

To conclude, we have systematically investigated the hydrogen-induced magnetic response in Co-Pd alloys and ultrathin Co/Pd multilayers through room-temperature characterization of the extraordinary Hall effect. Our study reveals that hydrogen does not produce a uniform magnetic response; instead, it shows a clear crossover between hysteresis contraction and expansion, governed by the competition between the electronic suppression of Pd magnetism and magnetoelastic enhancement of perpendicular anisotropy. In Pd-rich compositions, hydrogen fills the Pd-4d band and suppresses the induced Pd magnetic moment, resulting in the loop contraction and magnetic softening. With increasing effective Co fraction, the hydrogen-induced anisotropic lattice expansion produces an in-plane compressive stress, which in turn induces PMA through magnetoelastic coupling, resulting in loop expansion with increased coercivity and squareness. In interface-dominated ultrathin multilayers (as in the case of Co(0.1 nm) series), the partial compensation between the electronic and magnetoelastic effects results in a small net enhancement of coercivity and squareness, rather than a pronounced hardening or softening of hysteresis. EHE scaling analysis reveals a non-monotonic $\Delta\rho_h$–$\Delta\rho$ relationship, indicating that the EHE arises from multiple intrinsic and extrinsic contributions that are non-uniformly modified by hydrogen absorption. Hydrogen also induces EHE polarity reversal near the crossover region, suggesting an additional degree of tunability arising from changes in the dominant Hall scattering mechanisms. Together, these results establish a framework of composition-dependent magnetic response in the Co-Pd system and provide insights for engineering hydrogen-responsive spintronic devices through effective tuning of composition, thickness, and interface.

**Acknowledgements**

This research was supported by the State of Israel Ministry of Science, Technology, and Space (Grant No. 53453) and by the Israel Science Foundation (Grant No. 992/17).

**References**

[1] Y. Fukai, The Metal-Hydrogen System: Basic Bulk Properties, Springer, Berlin, Heidelberg, 2005. https://doi.org/10.1007/3-540-28883-X.

[2] F.A. Lewis, The Hydrides of Palladium and Palladium Alloys, Platin. Met. Rev. 4 (1960) 132–137.

[3] A. Pundt, R. Kirchheim, HYDROGEN IN METALS: Microstructural Aspects, Annu. Rev. Mater. Res. 36 (2006) 555–608. https://doi.org/10.1146/annurev.matsci.36.090804.094451.

[4] T.B. Flanagan, W.A. Oates, The Palladium-Hydrogen System, Annu. Rev. Mater. Sci. 21 (1991) 269–304. https://doi.org/10.1146/annurev.ms.21.080191.001413.

[5] G. Alefeld, J. Völkl, Hydrogen in Metals I: Basic Properties, Springer, Berlin, Heidelberg, 1978.

[6] B.D. Adams, A. Chen, The role of palladium in a hydrogen economy, Mater. Today 14 (2011) 282–289. https://doi.org/10.1016/S1369-7021(11)70143-2.

[7] F.D. Manchester, A. San-Martin, J.M. Pitre, The H-Pd (hydrogen-palladium) System, JPE 15 (1994) 62–83. https://doi.org/10.1007/BF02667685.

[8] T. Ozawa, H. Nakanishi, K. Kato, R. Shimizu, T. Hitosugi, K. Fukutani, Observation of resonant tunneling of proton from octahedral to tetrahedral sites in Pd, J. Phys. Chem. Solids 185 (2024) 111741. https://doi.org/10.1016/j.jpcs.2023.111741.

[9] J.P. Burger, S. Senoussi, B. Soufaché, Electrical and magnetic properties of palladium hydrides compared with those of pure palladium, J. Less-Common Met. 49 (1976) 213–222. https://doi.org/10.1016/0022-5088(76)90036-9.

[10] I.S. Maksymov, M. Kostylev, Magneto-Electronic Hydrogen Gas Sensors: A Critical Review, Chemosensors 10 (2022) 49. https://doi.org/10.3390/chemosensors10020049.

[11] S.S. Das, G. Kopnov, A. Gerber, Detection of hydrogen by the extraordinary Hall effect in CoPd alloys, J. Appl. Phys. 124 (2018) 104502. https://doi.org/10.1063/1.5049647.

[12] Y.-R. Chu, X.-W. Lu, C.-T. Hsieh, C.-Y. Huang, P.-H. Hsu, L.-J. Liaw, C.-M. Liu, W.-C. Lin, Hydrogen-mediated control of magnetic anisotropy and magnetic domain structure in Co/Pd multilayer, Appl. Phys. Lett. 126 (2025) 012404. https://doi.org/10.1063/5.0242410.

[13] G.L. Causer, M. Kostylev, D.L. Cortie, C. Lueng, S.J. Callori, X.L. Wang, F. Klose, In Operando Study of the Hydrogen-Induced Switching of Magnetic Anisotropy at the Co/Pd Interface for Magnetic Hydrogen Gas Sensing, ACS Appl. Mater. Interfaces 11 (2019) 35420–35428. https://doi.org/10.1021/acsami.9b10535.

[14] A. Gerber, G. Kopnov, M. Karpovski, Hall effect spintronics for gas detection, Appl. Phys. Lett. 111 (2017) 143505. https://doi.org/10.1063/1.4985241.

[15] C. Lueng, P.J. Metaxas, M. Sushruth, M. Kostylev, Adjustable sensitivity for hydrogen gas sensing using perpendicular-to-plane ferromagnetic resonance in Pd/Co Bi-layer films, Int. J. Hydrogen Energy 42 (2017) 3407–3414. https://doi.org/10.1016/j.ijhydene.2016.09.204.

[16] P.-C. Chang, C.-M. Liu, C.-C. Hsu, W.-C. Lin, Hydrogen-mediated magnetic domain formation and domain wall motion in $Co_{30}Pd_{70}$ alloy films, Sci. Rep. 8 (2018) 6656. https://doi.org/10.1038/s41598-018-25114-3.

[17] P.-C. Chang, Y.-Y. Chang, W.-H. Wang, F.-Y. Lo, W.-C. Lin, Visualizing hydrogen diffusion in magnetic film through magneto-optical Kerr effect, Commun. Chem. 2 (2019) 89. https://doi.org/10.1038/s42004-019-0189-1.

[18] K.-J. Hsueh, P.-C. Chang, L.-J. Liaw, A. Dhanarajagopal, M.-T. Lin, W.-C. Lin, Hydrogen-Controlled Spin Reorientation Transition in a Nanometer-Thick FePd Layer on Co/[Pt/Co]4/Pt Multilayers for Applications in Spintronics, ACS Appl. Nano Mater. 6 (2023) 2784–2790. https://doi.org/10.1021/acsanm.2c05095.

[19] S. Akamaru, N. Godo, S. Koshimoto, Magnetoresistance in Pd–Co/Cu/Pd–Co trilayer under hydrogen–nitrogen gas mixture, AIP Adv. 13 (2023) 095119. https://doi.org/10.1063/5.0161802.

[20] P.F. Carcia, A.D. Meinhaldt, A. Suna, Perpendicular magnetic anisotropy in Pd/Co thin film layered structures, Appl. Phys. Lett. 47 (1985) 178–180. https://doi.org/10.1063/1.96254.

[21] S. Hashimoto, Y. Ochiai, K. Aso, Perpendicular magnetic anisotropy and magnetostriction of sputtered Co/Pd and Co/Pt multilayered films, J. Appl. Phys. 66 (1989) 4909–4916. https://doi.org/10.1063/1.343760.

[22] G.H.O. Daalderop, P.J. Kelly, M.F.H. Schuurmans, Magnetic anisotropy of a free-standing Co monolayer and of multilayers which contain Co monolayers, Phys. Rev. B 50 (1994) 9989–10003. https://doi.org/10.1103/PhysRevB.50.9989.

[23] M.T. Johnson, P.J.H. Bloemen, F.J.A. den Broeder, J.J. de Vries, Magnetic anisotropy in metallic multilayers, Rep. Prog. Phys. 59 (1996) 1409. https://doi.org/10.1088/0034-4885/59/11/002.

[24] P. Bruno, Tight-binding approach to the orbital magnetic moment and magnetocrystalline anisotropy of transition-metal monolayers, Phys. Rev. B 39 (1989) 865–868. https://doi.org/10.1103/PhysRevB.39.865.

[25] D. Weller, Y. Wu, J. Stöhr, M.G. Samant, B.D. Hermsmeier, C. Chappert, Orbital magnetic moments of Co in multilayers with perpendicular magnetic anisotropy, Phys. Rev. B 49 (1994) 12888–12896. https://doi.org/10.1103/PhysRevB.49.12888.

[26] J. Okabayashi, Y. Miura, H. Munekata, Anatomy of interfacial spin-orbit coupling in Co/Pd multilayers using X-ray magnetic circular dichroism and first-principles calculations, Sci. Rep. 8 (2018) 8303. https://doi.org/10.1038/s41598-018-26195-w.

[27] T. Tokunaga, M. Kohri, H. Kadomatsu, H. Fujiwara, Magnetostriction of Pd–Co Alloys, J. Phys. Soc. Jpn. 50 (1981) 1411–1412. https://doi.org/10.1143/JPSJ.50.1411.

[28] S.U. Jen, B.L. Chao, Magnetostriction of polycrystalline Co-Pd alloys, J. Appl. Phys. 75 (1994) 5667–5669. https://doi.org/10.1063/1.355631.

[29] J. Carrey, A.E. Berkowitz, W.F. Egelhoff Jr., D.J. Smith, Influence of interface alloying on the magnetic properties of Co/Pd multilayers, Appl. Phys. Lett. 83 (2003) 5259–5261. https://doi.org/10.1063/1.1635660.

[30] P.F. Carcia, Perpendicular magnetic anisotropy in Pd/Co and Pt/Co thin-film layered structures, J. Appl. Phys. 63 (1988) 5066–5073. https://doi.org/10.1063/1.340404.

[31] F.J.A. den Broeder, W. Hoving, P.J.H. Bloemen, Magnetic anisotropy of multilayers, J. Magn. Magn. Mater. 93 (1991) 562–570. https://doi.org/10.1016/0304-8853(91)90404-X.

[32] K. Klyukin, G.S.D. Beach, B. Yildiz, Hydrogen tunes magnetic anisotropy by affecting local hybridization at the interface of a ferromagnet with nonmagnetic metals, Phys. Rev. Mater. 4 (2020) 104416. https://doi.org/10.1103/PhysRevMaterials.4.104416.

[33] G.H.O. Daalderop, P.J. Kelly, M.F.H. Schuurmans, First-principles calculation of the magnetic anisotropy energy of $Co_n/X_m$ multilayers, Phys. Rev. B 42 (1990) 7270–7273. https://doi.org/10.1103/PhysRevB.42.7270.

[34] S.-K. Kim, Y.-M. Koo, V.A. Chernov, J.B. Kortright, S.-C. Shin, Comparison of atomic structure anisotropy between Co-Pd alloys and Co/Pd multilayer films, Phys. Rev. B 62 (2000) 3025–3028. https://doi.org/10.1103/PhysRevB.62.3025.

[35] E. Wicke, H. Brodowsky, H. Züchner, Hydrogen in palladium and palladium alloys, in: G. Alefeld, J. Völkl (Eds.), Hydrogen in Metals II: Application-Oriented Properties, Springer, Berlin, Heidelberg, 1978: pp. 73–155. https://doi.org/10.1007/3-540-08883-0_19.

[36] A. Houari, S.F. Matar, V. Eyert, Electronic structure and crystal phase stability of palladium hydrides, J. Appl. Phys. 116 (2014) 173706. https://doi.org/10.1063/1.4901004.

[37] C. Lueng, F. Zighem, D. Faurie, M. Kostylev, Ferromagnetic resonance investigation of physical origins of modification of the perpendicular magnetic anisotropy in Pd/Co layered films in the presence of hydrogen gas, J. Appl. Phys. 122 (2017) 163901. https://doi.org/10.1063/1.4996808.

[38] S.S. Das, G. Kopnov, A. Gerber, Positive vs negative resistance response to hydrogenation in palladium and its alloys, AIP Adv. 10 (2020) 065129. https://doi.org/10.1063/5.0009194.

[39] S.S. Das, G. Kopnov, A. Gerber, Resistivity Testing of Palladium Dilution Limits in CoPd Alloys for Hydrogen Storage, Materials 15 (2022) 111. https://doi.org/10.3390/ma15010111.

[40] S. Akamaru, A. Kimura, M. Hara, K. Nishimura, T. Abe, Hydrogenation effect on magnetic properties of Pd–Co alloys, J. Magn. Magn. Mater. 484 (2019) 8–13. https://doi.org/10.1016/j.jmmm.2019.03.121.

[41] W.-C. Lin, B.-Y. Wang, H.-Y. Huang, C.-J. Tsai, V.R. Mudinepalli, Hydrogen absorption-induced reversible change in magnetic properties of Co–Pd alloy films, J. Alloys Compd. 661 (2016) 20–26. https://doi.org/10.1016/j.jallcom.2015.11.144.

[42] K. Ishida, T. Nishizawa, The Co-Pd (Cobalt-Palladium) System, JPE 12 (1991) 83–87. https://doi.org/10.1007/BF02663680.

[43] C. Morgan, K. Schmalbuch, F. García-Sánchez, C.M. Schneider, C. Meyer, Structure and magnetization in CoPd thin films and nanocontacts, J. Magn. Magn. Mater. 325 (2013) 112–116. https://doi.org/10.1016/j.jmmm.2012.07.052.

[44] B.D. Cullity, S.R. Stock, Elements of X-ray Diffraction, third ed., Prentice Hall, Upper Saddle River, NJ, 2001.

[45] A. Züttel, Materials for hydrogen storage, Mater. Today 6 (2003) 24–33. https://doi.org/10.1016/S1369-7021(03)00922-2.

[46] J. Völkl, G. Alefeld, Diffusion of hydrogen in metals, in: G. Alefeld, J. Völkl (Eds.), Hydrogen in Metals I: Basic Properties, Springer Berlin Heidelberg, Berlin, Heidelberg, 1978: pp. 321–348. https://doi.org/10.1007/3540087052_51.

[47] S.S. Das, G. Kopnov, A. Gerber, Kinetics of the lattice response to hydrogen absorption in thin Pd and CoPd films, Molecules 25 (2020) 3597. https://doi.org/10.3390/molecules25163597.

[48] J.W. Taylor, J.A. Duffy, J. Poulter, A.M. Bebb, M.J. Cooper, J.E. McCarthy, D.N. Timms, J.B. Staunton, F. Itoh, H. Sakurai, B.L. Ahuja, Spin-polarized electron momentum density distributions in Pd1-xCox alloys, Phys. Rev. B 65 (2001) 024442. https://doi.org/10.1103/PhysRevB.65.024442.

[49] B.D. Cullity, C.D. Graham, Introduction to Magnetic Materials, John Wiley & Sons, 2011.

[50] P.-C. Chang, Y.-C. Chen, C.-C. Hsu, V.R. Mudinepalli, H.-C. Chiu, W.-C. Lin, Hydrogenation-induced reversible spin reorientation transition in Co50Pd50 alloy thin films, J. Alloys Compd. 710 (2017) 37–46. https://doi.org/10.1016/j.jallcom.2017.03.221.

[51] G. Winer, A. Segal, M. Karpovski, V. Shelukhin, A. Gerber, Probing Co/Pd interfacial alloying by the extraordinary Hall effect, J. Appl. Phys. 118 (2015) 173901. https://doi.org/10.1063/1.4935023.

[52] N. Nagaosa, J. Sinova, S. Onoda, A.H. MacDonald, N.P. Ong, Anomalous Hall effect, Rev. Mod. Phys. 82 (2010) 1539–1592. https://doi.org/10.1103/RevModPhys.82.1539.

[53] S. Onoda, N. Sugimoto, N. Nagaosa, Intrinsic Versus Extrinsic Anomalous Hall Effect in Ferromagnets, Phys. Rev. Lett. 97 (2006) 126602. https://doi.org/10.1103/PhysRevLett.97.126602.

**Hydrogen-induced sign reversal in magnetic hysteresis evolution of CoPd alloys and Co/Pd multilayers**

S. S. Das[1,*] and A. Gerber[2,*]

[1]Institute of Industrial Science, The University of Tokyo, 4-6-1 Komaba, Meguro-ku, Tokyo 153-8505, Japan

[2]Raymond and Beverly Sackler Faculty of Exact Sciences, School of Physics and Astronomy, Tel Aviv University, Ramat Aviv, 69978 Tel Aviv, Israel

# Supplementary File

[*]Corresponding authors: ssdas@iis.u-tokyo.ac.jp (S. S. Das) and gerber@tauex.tau.ac.il (A. Gerber)

**S1: EHE loop measurement sequence highlighting $H_2$ exposure time**

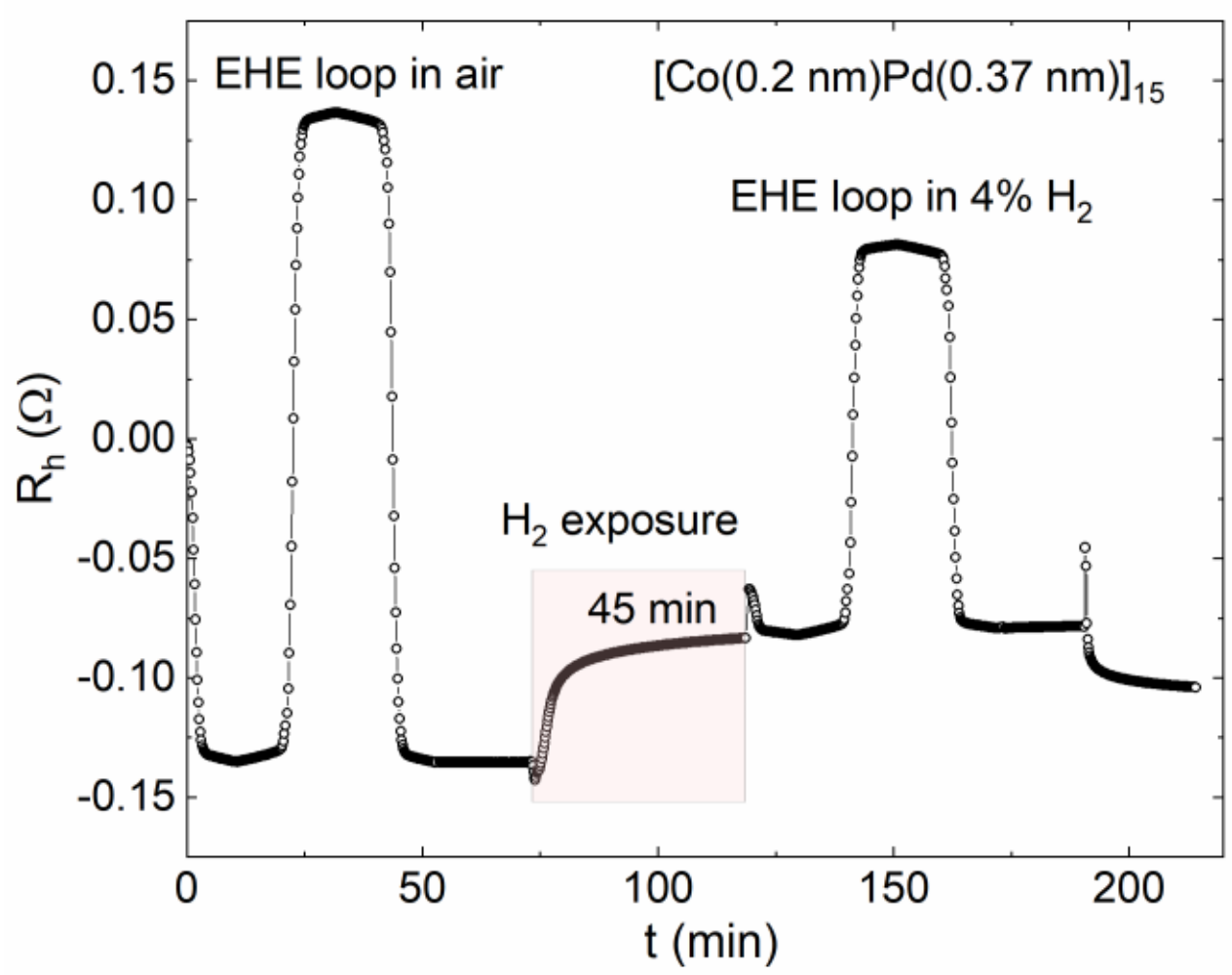

**Fig. S1** Extraordinary Hall resistance $R_h$ as a function of time for a representative [Co(0.2 nm)/Pd(0.37 nm)]$_{15}$ multilayer during a hydrogenation cycle. The initial and final loops correspond to measurements in air and after $H_2$ exposure, respectively. The shaded region (~ 45 min) indicates the $H_2$ exposure period until $R_h$ saturates, after which the EHE loop in $H_2$ is recorded.

**S2: Magnetic changes in alloys**

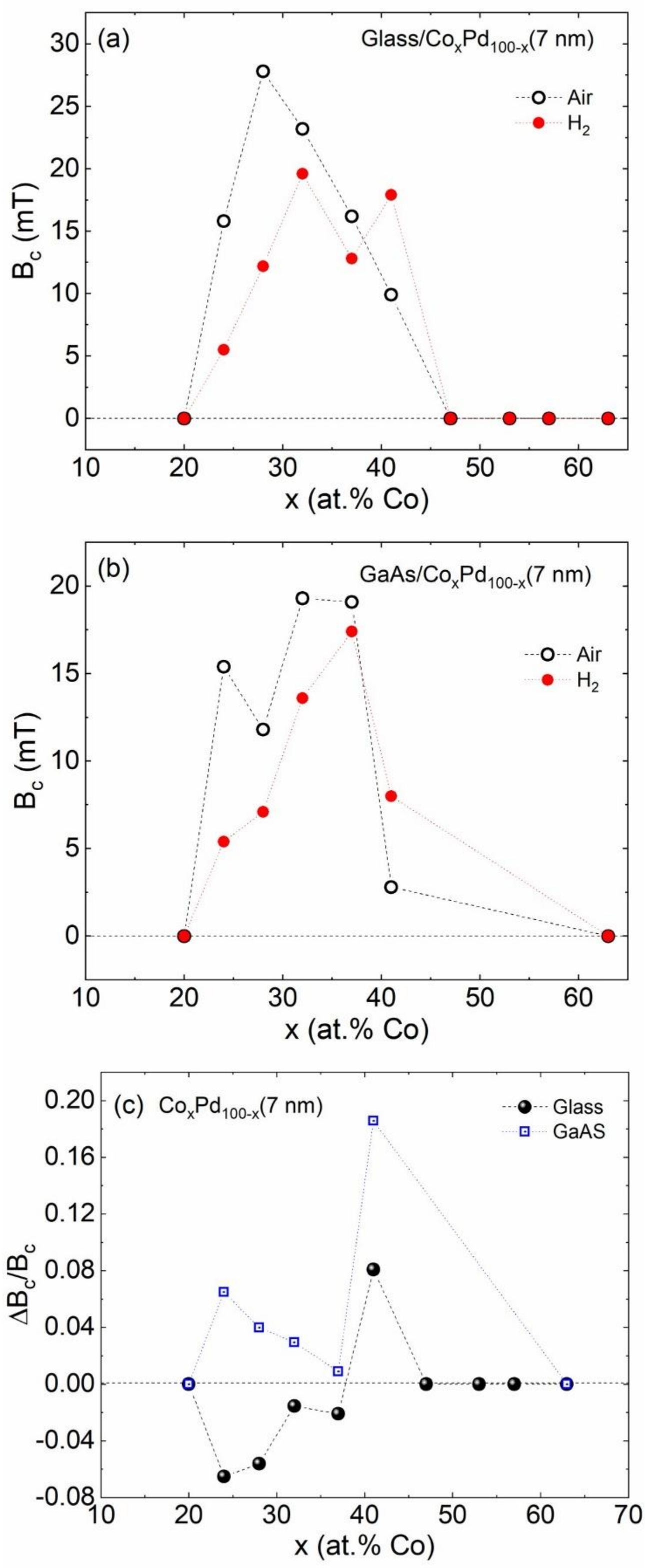

**Fig. S2** (a) Coercive field $B_c$ as a function of Co concentration x for glass/$Co_xPd_{100-x}$(7 nm) alloy films measured in air and under 4% $H_2$. (b) Corresponding $B_c$ variation for GaAs/$Co_xPd_{100-x}$(7 nm) films. (c)

Relative change in coercive field $\Delta B_c/B_c$ as a function of x for films grown on glass and GaAs substrates, where $\Delta B_c = B_c(H_2) – B_c(air)$. The dashed horizontal line indicates $\Delta B_c = 0$. Dashed lines are guides to the eye.

**S3: Magnetic changes in the multilayers**

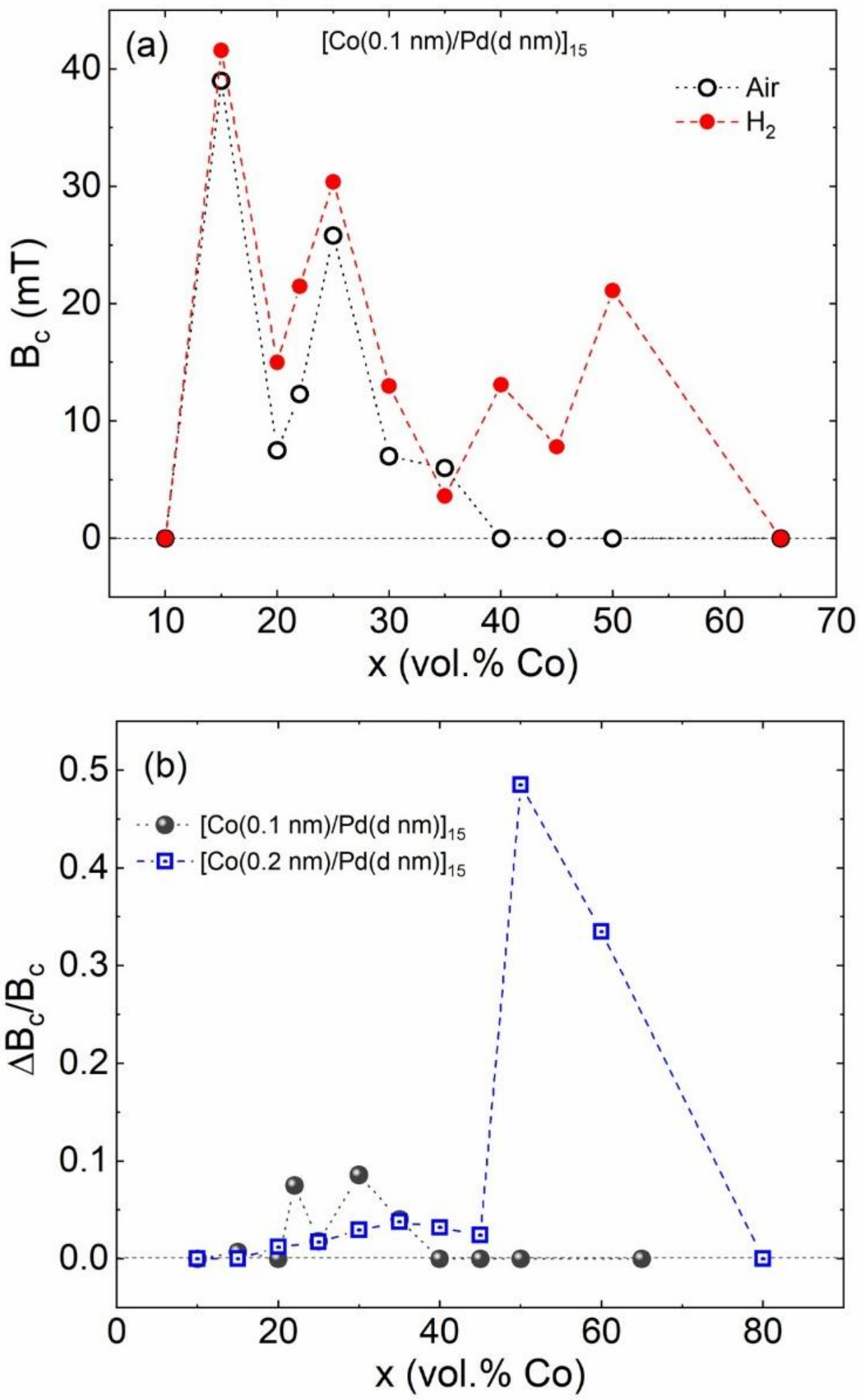

**Fig. S3**(a) Coercivity $B_c$ as a function of Co concentration x for the $[Co(0.1\ nm)/Pd(d)]_{15}$ multilayers measured in air and under 4% $H_2$. The data exhibit a non-monotonic dependence on composition, with hydrogen generally giving a positive shift in $B_c$ across most x. (b) Relative change in coercivity $\Delta B_c/B_c$ as a function of effective Co volume fraction x for both $[Co(0.1\ nm)/Pd(d)]_{15}$ and $[Co(0.2\ nm)/Pd(d)]_{15}$ multilayers.

**S4: EHE scaling with resistivity of $[Co(0.1\ nm)/Pd\ (d\ nm)]_{15}$ multilayer**

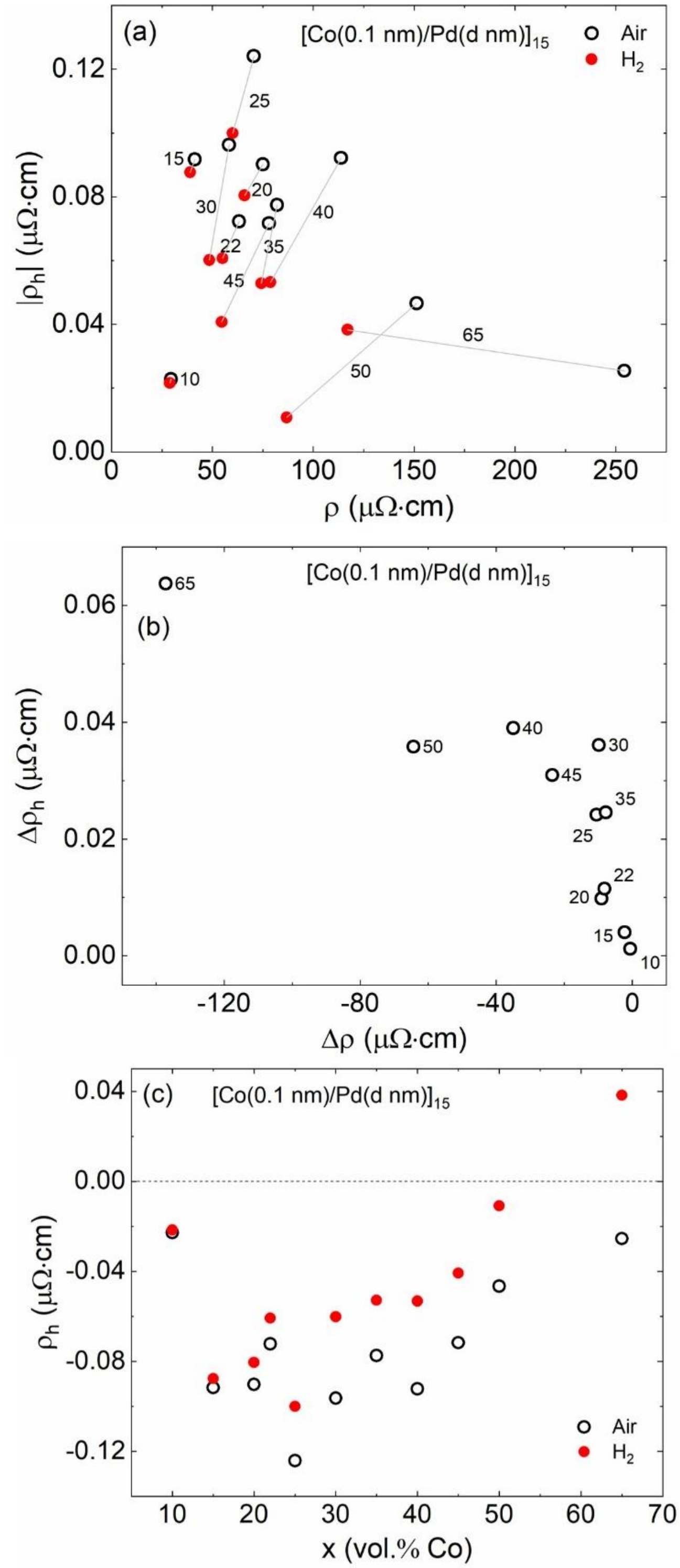

**Fig. S4** (a) Scaling of the extraordinary Hall (EHE) resistivity, $|\rho_h|$, as a function of the longitudinal resistivity, $\rho$, for $[Co(0.1\ nm)/Pd(d\ nm)]_{15}$ multilayers measured in air and under 4% $H_2$. The lines serve as guides to the eye, connecting corresponding air and hydrogen data points for each Co concentration.

(b) Hydrogen-induced changes in the Hall resistivity, $\Delta\rho_h$(= $\rho_h(H_2)$- $\rho_h$(air)), plotted as a function of the corresponding change in longitudinal resistivity, $\Delta\rho$ (= $\rho(H_2)$- $\rho$(air)), showing deviation from linear scaling. (c) $\rho_h$ as a function of x in air and $H_2$ for [Co(0.1 nm)/Pd(d nm)]$_{15}$ multilayers. The dashed horizontal line indicates $\rho_h = 0$.

**S5: Concentration dependence of EHE hysteresis evolution**

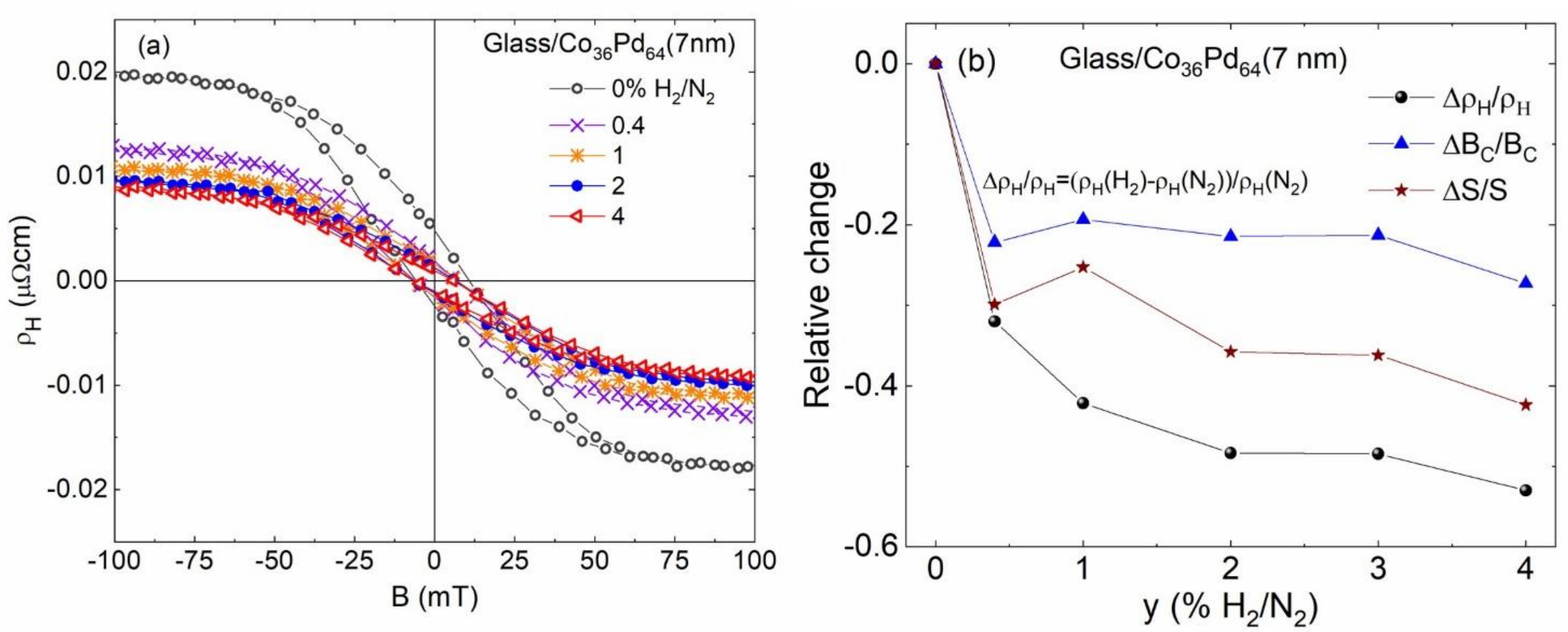

**Fig. S5** Representative $H_2$-concentration-dependent EHE response of Glass/$Co_{36}Pd_{64}$(7 nm) film. (a) EHE hysteresis loops measured under different $H_2/N_2$ concentrations (y) from 0 to 4%. (b) Relative changes in Hall resistivity, $\Delta\rho_H/\rho_H = [\rho_H(H_2) - \rho_H(N_2)]/\rho_H(N_2)$, and coercivity, $\Delta B_c/B_c$, and loop squareness, $\Delta S/S$, as a function of y. The data show the systematic evolution of EHE amplitude, coercive field, and loop squareness with increasing $H_2$ concentration. It also supports the use of 4% $H_2/N_2$ as a fixed near-saturated exposure condition for comparing the composition-dependent hydrogen response in the main text.